\documentclass[12pt]{article}

\usepackage[margin=1in]{geometry}

\usepackage{amsmath}
\usepackage{amsmath}
\usepackage{graphicx}
\usepackage{subcaption}
\usepackage{natbib}
\usepackage{url} 
\usepackage{xcolor}
\usepackage{csquotes}

\usepackage{
	amsmath,
	amssymb,
	mathtools,
	thmtools,
	bbm,
	subcaption,
	multirow,
	multicol
	}

\usepackage[utf8]{inputenc}
\usepackage{natbib}

\renewcommand{\appendixname}{}

\usepackage[capitalize,nameinlink]{cleveref}
\usepackage{crossreftools}

\crefname{section}{Section}{Sections}
\crefname{subsection}{Subsection}{Subsections}
\crefname{lemma}{Lemma}{Lemmas}
\crefname{corollary}{Corollary}{Corollaries}
\crefname{theorem}{Theorem}{Theorems}
\crefname{proposition}{Proposition}{Propositions}

\makeatletter
\renewcommand{\maketag@@@}[1]{\hbox{\m@th\normalsize\normalfont#1}}%
\makeatother

\makeatletter
\let\reftagform@=\tagform@
\def\tagform@#1{\maketag@@@{\ignorespaces\textcolor{gray}{(#1)}\unskip\@@italiccorr}}
\renewcommand{\eqref}[1]{\textup{\reftagform@{\ref{#1}}}}
\makeatother

\newcommand{\EE}{\mathbb{E}}

\newcommand{\NN}{\mathbb{N}}

\newcommand{\RR}{\mathbb{R}}

\newcommand{\zero}{0}

\def\[#1\]{\begin{equation}\begin{aligned}#1\end{aligned}\end{equation}}
\def\*[#1\]{\begin{equation*}\begin{aligned}#1\end{aligned}\end{equation*}}

\def\s*[#1\s]{\small\begin{align*}#1\end{align*}\normalsize}

\newcommand{\lcrx}[4][{-1}]{ 
	\IfEq{#1}{-1}{\left #2 {{{{#3}}}} \right #4}{
   	\IfEq{#1}{0}{#2 {{{{#3}}}} #4}{
	\IfEq{#1}{1}{\bigl #2 {{{{#3}}}} \bigr #4}{
	\IfEq{#1}{2}{\Bigl #2 {{{{#3}}}} \Bigr #4}{
	\IfEq{#1}{3}{\biggl #2 {{{{#3}}}} \biggr #4}{
	\IfEq{#1}{4}{\Biggl #2 {{{{#3}}}} \Biggr #4}{
    \GenericWarning{"4th argument to lcrx must be -1, 0, 1, 2, 3, or 4"}
    }}}}}}} 

\def\multiset#1#2{\ensuremath{\left(\kern-.3em\left(\genfrac{}{}{0pt}{}{#1}{#2}\right)\kern-.3em\right)}}

\DeclareMathOperator*{\argmin}{\arg\min} 

\DeclareMathOperator*{\newlim}{\mathrm{lim}\vphantom{\mathrm{infsup}}}
\DeclareMathOperator*{\newmin}{\mathrm{min}\vphantom{\mathrm{infsup}}}

\renewcommand{\lim}{\newlim}
\renewcommand{\min}{\newmin}

\newcommand{\Tr}{^{\scriptscriptstyle\text{T}}} 
\newcommand{\inv}{^{-1}}
\newcommand{\tr}{\mathrm{Tr}} 

\DeclareMathOperator{\diag}{diag} 

\newcommand{\grad}{\nabla} 

\newcommand{\normaldist}{\mathrm{N}}

\newcommand{\sbra}[2][{-1}]{\lcrx[#1] [ {#2} ] }

\newcommand{\norm}[1]{\lVert #1 \rVert}

\newcommand{\Nats}{\NN}

\newcommand{\Reals}{\RR}

\newcommand{\range}[2][{1}]{
	\IfEq{#1}{1}{\sbra{#2}}{\sbra{#2}_{#1}}}
\newcommand{\rangeO}[2][{0}]{
	\IfEq{#1}{0}{\sbra{#2}_0}{\sbra{#2}_{#1}}}

\newcommand{\mb}[1]{#1}

\newcommand{\regparamdim}{p}
\newcommand{\numcomponents}{d}
\newcommand{\componentdim}{m}
\newcommand{\varcompidx}{\tau}
\newcommand{\varcomp}{\mb{\varcompidx}}
\newcommand{\varcompconstr}{\widetilde{\mb{\varcompidx}}}
\newcommand{\varcompconstridx}{\widetilde{\varcompidx}}
\newcommand{\varcompmle}{\widehat{\varcomp}}

\newcommand{\varcompmlenull}{\varcompmle_0}

\newcommand{\varcompparamspace}{\mathcal{T}_\numobs}
\newcommand{\varcompparamspacezero}{\mathcal{T}_\infty}
\newcommand{\varcompparamspaceU}{\overline{\mathcal{T}}_n}
\newcommand{\lcmatrix}{\mb{A}}
\newcommand{\nrsidx}{q}
\newcommand{\nrs}{\mb{\nrsidx}}
\newcommand{\Nrs}{\mb{Q}}
\newcommand{\nrll}{\mathcal{L}}

\newcommand{\numboot}{B}
\newcommand{\nrsboot}{\nrs^{*}}

\newcommand{\sumsquare}[1]{\text{SS}_{#1}}

\newcommand{\varcompmleboot}{\varcompmle^{*}}
\newcommand{\pboot}{p_0^{*}}

\newcommand{\obs}{y}
\newcommand{\obsall}{\mb{\obs}}
\newcommand{\covi}{x}
\newcommand{\cov}{\mb{\covi}}

\newcommand{\reidx}{u}
\newcommand{\re}{\mb{\reidx}}

\newcommand{\regparami}{\beta}
\newcommand{\regparam}{\mb{\regparami}}

\newcommand{\indsim}{\overset{\text{\upshape \tiny ind.}}{\sim}}
\newcommand{\iidsim}{\overset{\text{\upshape \tiny iid.}}{\sim}}

\newcommand{\numobs}{N}

\begin{document}


\title{Testing Linear Combinations of Multiple Variance Components}
\author{Alex Stringer\thanks{
  The authors gratefully acknowledge funding from the Natural Sciences and Engineering Research Council of Canada}\hspace{.2cm}\\
  Department of Statistics and Actuarial Science, University of Waterloo\\
  and \\
  Jeffrey Negrea \\
  Department of Statistics and Actuarial Science, University of Waterloo}
\maketitle
\begin{abstract}
We test the hypothesis that simulataneous linear contrasts of multiple
variance components equal zero in a Gaussian variance components model via a parametric bootstrap.
Applications include but are not limited to nested and crossed designs.
The main technical contributions are a computationally efficient decomposition of the
normalized residual log-likelihood that does not 
require the variance components to be non-negative or variance design matrices to be positive semi-definite,
a modified Newton method for its minimization,
and a method for efficient optimization and sampling under the null hypothesis that certain linear combinations of variance components equal zero.
A special case of the proposed procedure is a test for multiple variance components simulataneously equalling zero, 
for which a likelihood ratio test was not previously available. 
However, the proposed procedure is significantly more general.
\end{abstract}

\noindent%
\vfill

\newpage


\section{Introduction}\label{sec:intro}

\subsection{Variance components models}\label{subsec:contributions}

Random processes with multiple components of variation are of central interest
in statistics. Testing whether a single component of variation is significantly 
different from zero in such models is a challenging problem that has been addressed 
via exact tests, asymptotic arguments, and parametric bootstrapping. To the
best of our knowledge there does not exist a test for the hypothesis that multiple
sources of variation are significantly different from \emph{each other}, or other such
hypotheses about linear combinations of variance components. 
In this paper we provide a test of the hypothesis that
a linear contrast of variance components is zero against both two- and one-sided
alternatives.
This turns out to be a technically demanding task leading to contributions to likelihood
computation, optimization, and sampling for variance components models.

The Gaussian variance components model is:
\begin{equation}\label{eqn:varcompmodel}
\obsall \sim \normaldist
\left\{  \mb{X\beta}, \sigma^2
\left( \mb{I}_\numobs + \sum_{j=1}^{d}\tau_j\mb{V}_j \right)
\right\},
\end{equation}
Here $\obsall$ is an $\numobs$-dimensional observed continuous response and $\mb{X}$ is a $\numobs\times\regparamdim$-dimensional design matrix for the mean. 
The $\mb{V}_j$ are fixed, known $\numobs\times \numobs$ matrices having rank $m_j\ll \numobs$.
It is usually assumed (e.g. \citealt{battey2024anomaly,hui2019testing}) that $\tau_j\geq0$ and each individual $\mb{V}_j$ is positive semi-definite, and
under these assumptions the model may be constructed by specifying an $\numobs\times m_j$ full-rank matrix $\mb{Z}_j$ and to set $\mb{V}_j = \mb{Z}_j\mb{Z}_j\Tr$.
In the sequel, we construct our models in this manner but then do not assume that $\mb{V}_j$ is semi-definite when optimizing the likelihood,
for reasons that will become clear (\cref{subsec:rotated}).

The unknown parameters $\varcomp = (\varcompidx_1,\ldots,\varcompidx_\numcomponents)\Tr$ are called \emph{variance components}.
Following \citet{battey2024anomaly}, the regression coefficients $\regparam\in\Reals^\regparamdim$ and residual variance $\sigma^2>0$ are treated as nuisance parameters and inference about $\varcomp$ is based on an appropriate maximal invariant statistic (\cref{subsec:residuallikelihood}).

Let $\lcmatrix\in\Reals^{d_0\times d}$ have full row rank $d_0 \leq d$.
We propose a parametric bootstrap residual log-likelihood ratio test of the linear hypothesis
\begin{equation}\label{eqn:test}
H_0: \lcmatrix\varcomp=\zero
\end{equation}
against a ``two-sided'' alternative $\lcmatrix\varcomp\neq\zero$, and more generally against the ``one-sided'' alternative $\lcmatrix\varcomp \in \mathcal{C}$ for some $\mathcal{C}\subset\Reals^{\numcomponents_0}$.
Examples include testing equality of variance components, $H_0:\tau_1=\cdots=\tau_{\numcomponents_0}, 1<\numcomponents_0\leq \numcomponents$, as well as testing that multiple components are null, $H_0: \tau_1=\cdots=\tau_{\numcomponents_0} = 0, 1\leq \numcomponents_0\leq \numcomponents$.
Prior work does not consider a general $\lcmatrix$ and instead tests the single-component ($\numcomponents_0=1$)
hypothesis $H_0:\varcompidx_d=0$ in a single-component model with $\numcomponents=1$ \citep{crainiceanu2004likelihood} or 
multiple-component model with $\numcomponents\geq 1$ \citep{wang2012testing,greven2008restricted,wood2013simple}.

The maximal parameter space of $\varcomp$ for which \cref{eqn:varcompmodel} defines a valid probability model and hence likelihood is a parametrization of the realtive interior of a \emph{spectrahedron}, an intersection of the positive semidefinite cone with an affine subspace, given by:
\begin{equation}\label{eqn:parameterspace}
\varcompparamspace = \left\{\varcomp\in\Reals^\numcomponents: \mb{I}_\numobs + \sum_{j=1}^{d}\tau_j\mb{V}_j \succ 0\right\}.
\end{equation}
Under the restriction that $\varcomp\in\varcompparamspacezero=[0,\infty)^\numcomponents$,
construction of \cref{eqn:varcompmodel} follows from the linear mixed model (\cref{eqn:randomeffectsmodel} in \cref{subsec:randomeffectsmodels})
where $\tau_j$ represents the variance of a latent variable, and computations rely upon the extensive machinery available for linear mixed models \citep{bates2015fitting}.
This restriction also arises naturally when considering asymptotic arguments
where $\numobs\to\infty$ \citep{stram1994variance}.
Methodological \citep{patterson1971recovery} and computational \citep{bates2015fitting} frameworks for variance component models tend to rely fundamentally on either $\varcompidx_j\neq0$
or $\varcompidx_j\geq0$, $j=1,\ldots,\numcomponents$, 
by assuming that certain matrices involving $\varcomp$ are invertible or have non-negative symmetric factorizations.
 
Our proposed test relies on a rotation of the model \cref{eqn:varcompmodel} given by \cref{eqn:varcompmodelrotated} in \cref{subsec:rotated}.
This will in turn require allowing for the case that some or all of the $\mb{V}_j$
are \emph{indefinite}, as long as $\mb{I}_\numobs + \tau_1\mb{V}_1 + \cdots + \tau_\numcomponents\mb{V}_\numcomponents \succ 0$.
A consequence of this relaxation of assumptions is that it is no longer 
meaningful to attribute a sign to individual $\tau_j$. 
Instead, we use the unrestricted parameter space $\varcompparamspace$ which includes $\varcomp=\zero$
as an interior point and allows for some $\tau_j<0$.
A convenient implication of this change in perspective is that we
avoid issues associated with the null hypothesis lying on the boundary
of the parameter space that plague testing hypotheses about variance components, e.g. in the work of \citet{crainiceanu2004likelihood,crainiceanu2005exact,wang2012testing,battey2024anomaly,stram1994variance,giampaoli2009likelihood,claeskens2004restricted}.

An inconvenient consequence of this change in perspective is that existing derivations of expressions for the (profiled) residual log-likelihood \citep{bates2015fitting,patterson1971recovery,crainiceanu2004likelihood} do not
apply, as their derivations fundamentally and irrevocably assume that $\varcompidx_j\geq0$ or $\varcompidx_j\neq0$ for each $j=1,\ldots,\numcomponents$.
\citet{battey2024anomaly} provide a derivation of the residual log-likelihood as the marginal distribution of a maximal invariant statistic for $\varcomp$
that does not require $\varcompidx_j\geq0$.
The resulting expression gives (up to additive constants) the same numerical value as that provided by \citet{bates2015fitting}
but, unlike \citet{bates2015fitting}, is not scalable in large models with sparse dependence structures as it involves decomposition and inversion of dense $O(\numobs)\times O(\numobs)$ matrices.
Together with the more complicated spectrahedral geometry of the maximal parameter space, $\varcompparamspace$, this leads to computational challenges
in likelihood computation and optimization that we address in \cref{sec:decomposition,sec:newton}.
The result is a computational framework for variance component models that
is efficient and accommodates indefinite $\mb{V}_j$ and negative $\varcompidx_j$, allowing us to proceed with our proposed parametric bootstrap
likelihood ratio test of $\lcmatrix\varcomp=\zero$.

We focus on parametric bootstrap likelihood ratio tests in this paper.
For testing the simpler multi-component hypothesis that $\tau_1=\cdots=\tau_{\numcomponents_0} = 0, \numcomponents_0 \leq \numcomponents$, the classic $F$-test is exact and easy to implement without fitting the model under the null or alternative hypothesis \citep{hui2019testing}.
However, allowing indefinite $\mb{V}_j$ as required for testing $\lcmatrix\varcomp=\zero$ means that the projections upon which the $F$-test
relies are not guaranteed to exist.
Together with the arguments of \citet{battey2024anomaly} favouring
likelihood ratio tests over Wald and score tests \citep{verbeke2003use,qu2013linear,zhang2024fast} for variance components, this
motivates our focus on likelihood ratio tests.
It remains to consider the use of approximate pivots based on asymptotic arguments, or exact (up to Monte Carlo error) tests based on bootstrapping.
\citet{crainiceanu2004likelihood} give an extensive argument in opposition of the use of asymptotic arguments for single component models $\numcomponents=1$ and hypotheses $\numcomponents_0=1$, favouring the parametric bootstrap. 
This argument is further supported by \citet{greven2008restricted,wood2013simple,wang2012testing} 
for multi-component models with single component hypotheses, $\numcomponents\geq1$ and $\numcomponents_0=1$. 
To our knowledge, no likelihood ratio test was previously available for general $1\leq \numcomponents_0\leq \numcomponents$ with $\numcomponents\geq1$,
which is a special case of our proposed test, but we see no reason why the arguments of these authors would not extend to this case.
Given this line of reasoning, and the efficient computational framework we develop, at present we favour bootstrapping.
Accordingly, we proceed with a parametric bootstrap likelihood ratio test.

\subsection{Linear mixed effects models}\label{subsec:randomeffectsmodels}

Variance components models are often obtained as the marginal distribution of the observable variables in the following
linear mixed effects model:
\[\label{eqn:randomeffectsmodel}
\obsall 
    & = \mb{X\beta} + \mb{Z}_1\re_1 + \cdots + \mb{Z}_\numcomponents\re_\numcomponents + \mb{\epsilon},\quad 
    & \re_j 
        & \indsim \normaldist\left(\zero, \sigma^2\varcompidx_j\mb{I}_{m_j}\right), \quad
    & \mb{\epsilon}
        &\sim\normaldist\left(\zero, \sigma^2\mb{I}_\numobs\right).
\]
The marginal distribution of $\mb{y}$ implied by \cref{eqn:randomeffectsmodel} is \cref{eqn:varcompmodel}.
The variance components are interpreted as the relative variance of each factor to the residual variance.
Considering $\varcompidx_j$ to be proportional to the variance of a latent random variable restricts the parameter space for $\varcomp$ from $\varcompparamspace$ to $\varcompparamspacezero$.
Though the linear mixed model conveniently expresses the modelled dependence structure in the data,
due to this artificial restriction of the parameter space, 
all mathematical and computational derivations in this
paper proceed directly from \cref{eqn:varcompmodel} with the parameter space $\varcompparamspace$.

Two specific types of variance component models to which our proposed test applies are named \emph{nested} and \emph{crossed} effects.
These are most conveniently expressed in linear mixed model form.
For example, three-level \emph{nested} random effects model has the form:
\[\label{eqn:nestedmodel}
\obs_{ijk} &= \cov_{ijk}\Tr\regparam + u_i + v_{ij} + \epsilon_{ijk}; & u_i & \iidsim \normaldist(0,\sigma^2\tau_1),& v_{ij}&\iidsim\normaldist(0,\sigma^2\tau_2), & \epsilon_{ijk}&\iidsim\normaldist(0,\sigma^2),
\]
where $i=1,\ldots,m$, $j = 1,\ldots,n_i$, and $k = 1,\ldots,r_{ij}$.
Such a nested model could be appropriate for blocked split-plot
designs such as the famous oat yield data of \citet{yates1935complex} (see \citet[Ch. 10]{venables2002random}).
Another example is a two-way \emph{crossed} random effects model of the form:
\[\label{eqn:crossedmodel}
\obs_{ijk} &= \cov_{ijk}\Tr\regparam + u_i + v_{j} + \epsilon_{ijk}; & u_i & \iidsim \normaldist(0,\sigma^2\tau_1),& v_{j}&\iidsim\normaldist(0,\sigma^2\tau_2), & \epsilon_{ijk}&\iidsim\normaldist(0,\sigma^2),
\]
where $i \in \{1,\ldots,m\}$, $j \in \{1,\ldots,n\}$, and $k \in \{1,\ldots,r_{ij}\}$.
Such a model could be used, for example, in an item-response study where the responses, $y_{ijk}$, represent the rating of the $j^{th}$ item by the $i^{th}$ subject, and $k=1$.
Each distinct pair $(i,j)$ may or may not be present in the data, and if present may have multiplicity greater than one.
When all pairs $(i,j)$ are present in the data, $m=n$, and $r_{ij}\equiv r$ is constant the design
is balanced, otherwise it is unbalanced.

\subsection{Motivating examples}\label{subsec:motivation}
In the absence of uncertainty quantification, the relative magnitudes of point estimates of variance components reported by statistical software 
are often used to draw conclusions which attribute variability in the response to specific sources of variation.
Statements of the form ``the variability due to factor A is twice as large as the variability due to factor B'' abound in the scientific literature, but are often based on point estimates of variance components without any statistical test.
\cite{hoffman2016variancepartition} provide a software package, \texttt{variancePartition}, for the purpose
of comparing components of variation, where 
all such comparisons are based on point estimation and ignore uncertainty quantification.
The aim of this work is to provide uncertainty quantification for such comparisons via a statistical test.
Accordingly, we describe examples from the literature in which such comparisons are described in the absence of uncertainty quantification.
Further details about all examples are included in the Supplementary Materials.

In the social sciences literature, \citet{nye2004large} study the variability of student outcomes with hierarchical dependence structure present through blocking
variables
``school'', and ``teacher'', 
as well as whether variation in these blocks
differ across fixed effects such as socioeconomic status. 
They draw conclusions that correspond to a null hypothesis of the form $H_0:\tau_1 = 2\tau_2$ and $\tau_3 = 3\tau_4$ against the
alternative that $\tau_1 > 2\tau_2$ or $\tau_3 > 3\tau_4$, but these conclusions are only based on
on point estimates of the variance components $\varcomp$ and the significance of a test that each $\tau_j\neq0$.
Our proposed test covers this setting with $\lcmatrix = ((1, -2, 0, 0), (0, 0, 1, -3))$, so $\numcomponents_0=2$ and $\numcomponents=3$.
In ecology, \citet{messier2010traits} examine the variation in traits across nested ecological scales including  ``site'', ``plot'', ``species'', ``tree'', ``strata'' and ``leaf''.
They make conclusions that correspond to a null hypothesis of the form $H_0: (\tau_1 = 0 \land \tau_2+\tau_3+\tau_4=\tau_5)$
against an alternative of the form $(\tau_1 \neq 0 \lor \tau_2+\tau_3+\tau_4 \neq \tau_5)$,
without performing such a test or considering uncertainty.
Again, our procedure covers this hypothesis with $\lcmatrix = ((1,0,0,0,0), (0,1,1,1,-1))$ having $\numcomponents_0=2$ and $\numcomponents=5$.
In genetics, \citet{hill2008data} partition the variability of complex traits three sources (additive, dominance, and epistatic),
and make conclusions corresponding to the null hypothesis $H_0:\tau_1 = \tau_2 + \tau_2$.
This can be formally tested using our method with $\lcmatrix = (1, -1, -1)$ having $\numcomponents_0=1$ and $\numcomponents=3$.

Examples from the statistics literature are abundant. 
\cref{tab:examples} shows results from three crossed and five nested designs for which variance component estimates were previously reported in statistical textbooks \citep{pinheiro2000mixed,venables2002random} or papers \citep{yates1935complex,immer1934barley,wright2013barley,davies1947statistical} 
and for which the data are presently available in the \textsf{R} language for statistical computing \citep{Rlanguage}.
Full descriptions and details are given in the Supplementary Materials, and a more thorough analysis of several of the data sets in \cref{tab:examples}
is given in \cref{sec:dataanalysis}.
The $p$-values shown in \cref{tab:examples}
are from application of our proposed test to these examples. 
In all cases an $F$-test rejects $H_0:\tau_j=0$ for $j=1,2$ at the $5\%$ level, but the hypothesis that $\tau_1=\tau_2$
is sometimes rejected and sometimes not, sometimes in apparent contrast to a very small or very large point estimate of the difference/ratio of variance components.
The analysis in \cref{sec:dataanalysis} reveals substantial variability in the bootstrapped sampling distributions of variance components across different designs and data sets,
leading to cases in which an apparently huge point estimate of the relative difference fails to be significantly different from zero at the (typical arbitrary) $5\%$
significance level.
This all underscores the need for a hypothesis test to be used when conclusions are to be made about the relative magnitudes of variance components.

\begin{table}
\centering
\begin{tabular}{llllll}
Name & Design & Sample size & $\varcompmle$ & $H_1:\tau_1 \neq \tau_2$ & $H_1:\tau_1 > \tau_2$  \\
\hline
Pastes & Nested & $60=10\times2\times3$ & $(12.49, 2.44)$ & $0.171\pm0.024$ & $0.13\pm0.021$ \\
Oats & Nested & $72=6\times3\times4$ & $(1.32, 0.675)$ & $0.57\pm0.03$ & $0.18\pm0.024$ \\
Machines & Crossed & $54=3\times6\times3$ & $(4.82, 2.65)$ & $0.64\pm0.03$ & $0.25\pm0.027$ \\
Penicillin & Crossed & $144=24\times6$ & $(12.34, 2.37)$ & $\approx0$ & $\approx0$ \\
Alfalfa & Nested & $72=6\times4\times3$ & $(1.16, 0.6)$ & $0.56\pm0.0314$ & $0.15\pm0.0227$ \\
Barley & Crossed & $60=6\times10$ & $(4.85, 0.56)$ & $0.012\pm0.0069$ & $0.004\pm0.004$ \\
Oxide & Nested & $72=8\times3\times3$ & $(10.34, 2.85)$ & $0.091\pm0.018$ & $0.019\pm0.0086$ \\
\end{tabular}
\caption{Some datasets available in \texttt{R} with continuous response and $d=2$ variance components
in a crossed or nested design. 
In all examples an F-test of $H_0:\tau_j=0$ for both $j=1,2$ has a $p$-value of $\approx0$.
In some examples the hypothesis of $\tau_1 = \tau_2$ is rejected and some not by our new test, despite apparently
large or small point estimated differences in variance components.}
\label{tab:examples}
\end{table}


\section{Testing the linear hypothesis}\label{sec:prelims}

\subsection{Rotated variance components model}\label{subsec:rotated}

Let $\lcmatrix\Tr = \mb{Q}_A(\mb{R}_A\Tr,\zero\Tr)\Tr$ where
$\mb{Q}_A = [\mb{Q}_1:\mb{Q}_2]$ is $\numcomponents\times\numcomponents$ orthonormal with $\text{dim}(\mb{Q}_1) = \numcomponents\times \numcomponents_0$ and $\text{dim}(\mb{Q}_2) = \numcomponents\times(\numcomponents-\numcomponents_0)$.
Let $\varcompconstr = \mb{Q}_A\Tr\varcomp = (\varcompconstr_1\Tr,\varcompconstr_2\Tr)\Tr$ where $\varcompconstr_1 = \mb{Q}_1\Tr\varcomp$ and $\varcompconstr_2 = \mb{Q}_2\Tr\varcomp$. 
Note that $\varcomp = \mb{Q}_A(\varcompconstr_1\Tr,\varcompconstr_2\Tr)\Tr$ and $\lcmatrix\varcomp=\zero$ if and only if $\varcompconstr_1=\zero$.
For $j = 1,\ldots,d$ let 
$$
\widetilde{\mb{V}}_j = \sum_{k=1}^{d}(\mb{Q}_A)_{kj}\mb{V}_k.
$$
Writing the variance components model in rotated form,
\begin{equation}\label{eqn:varcompmodelrotated}
\obsall \sim \normaldist
\left\{  \mb{X\beta}, \sigma^2
\left( \mb{I}_\numobs + \varcompconstridx_1\widetilde{\mb{V}}_1 + \cdots + \varcompconstridx_\numcomponents\widetilde{\mb{V}}_\numcomponents \right)
\right\},
\end{equation}
we see that $H_0:\lcmatrix\varcomp=\zero$ is equivalent to $H_0: \varcompconstr_1 = \zero$.
The rotated variance component matrices $\widetilde{\mb{V}}_j$ are linear combinations of the semi-definite matrices $\mb{V}_k$
with weights from the entries of the orthonormal matrix $\mb{Q}_A$. If some of these weights are possibly negative then 
$\widetilde{\mb{V}}_j$ may be \emph{indefinite} in general.
Recall that every orthonormal matrix is either a permutation or contains a negative entry.
If $\lcmatrix$ is a permutation of a diagonal matrix then testing $H_0:\lcmatrix\varcomp=\zero$ is equivalent to testing
$H_0:\tau_1=\cdots=\tau_{\numcomponents_0}=0$ for some $1\leq \numcomponents_0\leq d$ and can already be achieved using an $F$-test when the $\mb{V}_i$ are positive semi-definite.
It follows that $\mb{Q}_A$ must be allowed to contain at least one negative entry in order for the proposed test to apply to any
examples that cannot already be tested. Hence in the remainder we assume that the $\widetilde{\mb{V}}_j$ may be indefinite.

A parametric bootstrap test of $H_0:\lcmatrix\varcomp=\zero$
will require sampling from \cref{eqn:varcompmodelrotated} under $H_0:\varcompconstr_1=\zero$
(\cref{sec:bootstrap}) and fitting \cref{eqn:varcompmodelrotated} with potentially indefinite $\widetilde{\mb{V}}_j$.
A fundamental implication is that the sign of $\varcompconstridx_j$ is no longer meaningful, and restricting to $\varcompconstridx_j\geq0$ is arbitrary.
While the underlying random effects model is still assumed to be constructed with $\mb{V}_j=\mb{Z}_j\mb{Z}_j\Tr$, 
the likelihood optimization (\cref{sec:newton}) and bootstrap sampling (\cref{sec:bootstrap}) will not assume that $\widetilde{\mb{V}}_j$ is positive semi-definite or that $\varcompidx_j\geq0$. The result will be the development of a computational framework for fitting model (\ref{eqn:varcompmodel}) with indefinite $\mb{V}_j$ which will then be applied to (\ref{eqn:varcompmodelrotated}) to test $H_0:\lcmatrix\varcomp=\zero$.

\subsection{Normalized residual likelihood}\label{subsec:residuallikelihood}

Write $\varcompidx_1\mb{Z}_1\mb{Z}_1\Tr + \cdots + \varcompidx_d\mb{Z}_d\mb{Z}_d\Tr = \mb{Z}\mb{D}(\varcomp)\mb{Z}\Tr$
with $\mb{Z}=[\mb{Z}_1:\cdots:\mb{Z}_d]$ and $\mb{D}(\varcomp)$ diagonal.
It follows that
\begin{equation}\label{eqn:covmat}
\mb{\Sigma}(\varcomp) = \mb{I}_N + \mb{Z}\mb{D}(\varcomp)\mb{Z}\Tr.
\end{equation}
Let $\mb{U}_X\in\Reals^{\numobs\times(\numobs-p)}$ be an orthonormal
basis for $\text{ker}(\mb{X})$ so that $\mb{U}_X\Tr\mb{X}=\zero$.
Residual maximum likelihood \citep{patterson1971recovery} estimates $\sigma^2$
from the marginal likelihood of residuals $\mb{U}_X\Tr\obsall$ and $\varcomp$ from the marginal
likelihood of the normalized residuals $\mb{U}_X\Tr\obsall / \norm{\mb{U}_X\Tr\obsall}_2$.
The original approach of \citet{patterson1971recovery} for $d=1$ starts by assuming that $\mb{D}(\varcomp)\inv$ exists (Eq. (3) of \citealt{patterson1971recovery}) which
implies $\varcompidx_j\neq0$ for each $j=1,\ldots,\numcomponents$ and hence cannot be used directly as the basis of a test that $H_0:\varcomp=\zero$.
In a more general setup allowing for correlated random effects (e.g. intercepts and slopes),
\citet{bates2015fitting} instead assume that (their equivalent of) $\mb{D}(\varcomp)$ can be factored as $\mb{D}(\varcomp) = \mb{D}(\varcomp)^{1/2}\mb{D}(\varcomp)^{1/2}$ (e.g. Eq. (11) of \citet{bates2015fitting}) which assumes $\varcompidx_j\geq0$, $j=1,\ldots,\numcomponents$.
They give an expression for efficient computation of the profiled residual likelihood for $\varcomp$ based on this factorization.
This uses the ``restricted'' marginal likelihood approach of \citet{laird1982random} which is
obtained by integrating the marginal likelihood of $\obsall$ implied by \cref{eqn:randomeffectsmodel} over $\mb{\beta}$ and then eliminating $\sigma^2$ using profiled likelihood.
The assumption that $\varcompidx_j\geq0$ is also used explicitly in their optimization routines.

An alternative is to base inference on the marginal distribution of a maximal invariant statistic for $\varcomp$.
The \emph{normalized residual statistic} is
$
\Nrs = \mb{U}_X\Tr\obsall / \norm{\mb{U}_X\Tr\obsall}_2,
$
with observed value $\nrs\in\Reals^{\numobs-\regparamdim}$.
Twice the normalized residual negative log-likelihood 
obtained from the marginal density of $\Nrs$ is \citep{battey2024anomaly}:
\begin{equation}\label{eqn:nrll}
\nrll(\varcomp) = \log\lvert \mb{U}_X\Tr\mb{\Sigma}(\varcomp)\mb{U}_X\rvert + (\numobs-\regparamdim)\log\left[ \nrs\Tr \left\{\mb{U}_X\Tr\mb{\Sigma}(\varcomp)\mb{U}_X\right\}\inv \nrs \right].
\end{equation}
Evaluation of $\nrll(\varcomp)$ for a given $\varcomp$ requires decomposition of the $(\numobs-p)\times(\numobs-p)$-dimensional matrix
$\mb{U}_X\Tr\mb{\Sigma}(\varcomp)\mb{U}_X$. 
This matrix is dense even in the typical case that $\mb{Z}$ and hence $\mb{\Sigma}(\varcomp)$ is sparse.
The computational cost of evaluating Eq. (\ref{eqn:nrll}) is therefore $O(\numobs^3)$.
The profiled REML of \citet[Eq. (41)]{bates2015fitting} has the same numerical value as Eq. (\ref{eqn:nrll}) but requires only
decompositions of sparse matrices and is therefore computationally efficient. \citet{bates2015fitting} obtain their efficient expression
by assuming that $\varcompidx_j\geq0$. 
In the sequel we derive a decomposition of \cref{eqn:nrll} that leads to efficient computation of $\nrll(\varcomp)$ and is applicable
when the variance component matrices are indefinite and $\varcomp\in\varcompparamspace$.


\section{Decomposition of the Residual Likelihood}\label{sec:decomposition}

An efficient and stable method for fitting the model described by \cref{eqn:varcompmodel} by minimizing \cref{eqn:nrll} is required in
order for the parametric bootstrap to be feasible.
This in turn requires a computationally efficient representation of
$\nrll(\varcomp)$ that avoids decomposition and inversion of dense
$\numobs\times \numobs$-dimensional matrices.
We focus on the case in which $\mb{\Sigma}(\varcomp)$ is sparse.
This covers common linear mixed models for grouped data including nested and crossed designs.
Our approach allows for the variance components matrices to be indefinite as long as $\mb{\Sigma(\varcomp)}$ is strictly positive-definite.
The optimization in \cref{sec:newton} will ensure this.

We base our decomposition on orthogonal and triangular decompositions. 
These decompositions preserve sparsity when present,
and different algorithms are available for sparse or
dense matrices. 
For clarity we omit in our derivation the row/column permutations
that are a fundamental part of the algorithms for sparse orthogonal and
triangular decomposition. Our implementation uses \texttt{CHOLMOD} \citep{chen2008algorithm} interfaced through \texttt{julia} \citep{bezanson2017julia} and does use
these permutations.

Consider the orthogonal ($QR$) decompositions of the design matrices $\mb{X} = \mb{P}_X(\mb{R}_X\Tr,\zero\Tr)\Tr$ and $\mb{Z} = \mb{P}_Z(\mb{R}_Z\Tr,\zero\Tr)\Tr$
where $\mb{P}_X = [\mb{Q}_X:\mb{U}_X]$ with $\mb{Q}_X\in\Reals^{\numobs\times p}$ and $\mb{P}_Z = [\mb{Q}_Z:\mb{U}_Z]$ with $\mb{Q}_Z\in\Reals^{\numobs\times m}$.
The key advantage is that both dense and sparse algorithms are available for this, returning either dense or sparse $\mb{R}_Z$
factors as appropriate and exploiting sparsty of $\mb{Z}$ in computations when it is present. 
In both cases the orthogonal matrices $\mb{P}_X$ and $\mb{P}_Z$ are dense.
These matrices are never formed explicitly and are instead represented as a product
of (dense or sparse) Householder transformations with efficient algorithms available for computing matrix products and solving linear systems.

The original derivation of residual likelihood due to \citet{patterson1971recovery} for $\numcomponents=1$
applies the Woodbury identity in a manner that requires $\mb{D}(\varcomp) = \varcompidx_1\mb{I}_{m_1}$ be invertible and hence $\varcompidx_j\neq0$.
In a similar manner but avoiding this requirement, 
we apply block determinant and inverse identities to obtain
\begin{align}\label{eqn:blockdetinv}
\lvert\mb{U}_X\Tr\mb{\Sigma}(\varcomp)\mb{U}_X\rvert &= \lvert\mb{\Sigma}(\varcomp)\rvert\cdot\lvert\mb{Q}_X\Tr\mb{\Sigma}(\varcomp)\inv\mb{Q}_X\rvert, \\
\obsall\Tr\mb{U}_X\left\{\mb{U}_X\Tr\mb{\Sigma}(\varcomp)\mb{U}_X\right\}\inv\mb{U}_X\Tr\obsall &= \obsall\Tr\mb{\Sigma}(\varcomp)\inv\obsall \\
& \quad - \obsall\Tr\mb{\Sigma}(\varcomp)\inv\mb{Q}_X\left\{ \mb{Q}_X\Tr\mb{\Sigma}(\varcomp)\inv\mb{Q}_X\right\}\inv\mb{Q}_X\Tr\mb{\Sigma}(\varcomp)\inv\obsall.
\end{align}
To compute these quantities efficiently, let $\mb{L}(\varcomp)\mb{L}(\varcomp)\Tr = \mb{R}_Z\mb{D}(\varcomp)\mb{R}_Z\Tr + \mb{I}_m$ 
be the (dense or sparse) Cholesky decomposition and write
\[\label{eqn:sigmadecomp}
\mb{\Sigma}(\varcomp) & = \mb{P}_Z\begin{pmatrix} \mb{L}(\varcomp)\mb{L}(\varcomp)\Tr & 0 \\ 0 &  \mb{I}_{\numobs-m}\end{pmatrix}\mb{P}_Z\Tr, \\
\mb{\Sigma}(\varcomp)\inv &= \mb{P}_Z\begin{pmatrix} \{\mb{L}(\varcomp)\inv\}\Tr\mb{L}(\varcomp)\inv & 0\\ 0 & \mb{I}_{\numobs-m}\end{pmatrix}\mb{P}_Z\Tr.
\]
We then have
\begin{equation}\label{eqn:sigmainvidentities}
\begin{aligned}
\lvert\mb{\Sigma}(\varcomp)\rvert &= \lvert\mb{L}(\varcomp)\rvert^2 = \prod_{j=1}^m \{\mb{L}(\varcomp)_{jj}\}^2, \\
\mb{Q}_X\Tr\mb{\Sigma}(\varcomp)\inv\obsall &= \begin{pmatrix}\mb{L}(\varcomp)\inv(\mb{P}_Z\Tr\mb{Q}_X)_{1:m} \\ (\mb{P}_Z\Tr\mb{Q}_X)_{(m+1):\numobs}\end{pmatrix}\Tr\begin{pmatrix}\mb{L}(\varcomp)\inv(\mb{P}_Z\Tr\obsall)_{1:m} \\ (\mb{P}_Z\Tr\obsall)_{(m+1):\numobs}\end{pmatrix},
\end{aligned}
\end{equation}
with similar expressions available for the scalar $\obsall\Tr\mb{\Sigma}(\varcomp)\inv\obsall$ and $p\times p$ matrix $\mb{Q}_X\Tr\mb{\Sigma}(\varcomp)\inv\mb{Q}_X$.
The $N\times p$-dimensional matrix $\mb{Q}_X$ is formed explicitly and computations involving matrices conformable to $\mb{Q}_X\Tr\mb{Q}_X$ are 
performed naively as $p$ is assumed to be small.
The products $\mb{P}_Z\Tr\obsall$ and $\mb{P}_Z\Tr\mb{Q}_X$
are computed without forming $\mb{P}_Z$ using its Householder product
representation. The $m$-dimensional sparse triangular systems involving
$\mb{L}(\varcomp)$ are also solved efficiently. 
Efficient likelihood computation results from using \cref{eqn:sigmainvidentities} and its variants in \cref{eqn:nrll}.

\section{Minimizing $\nrll(\varcomp)$}\label{sec:newton}

\subsection{Modified Newton iteration}\label{subsec:modifednewton}

The negative residual log-likelihood $\nrll(\varcomp)$ is a twice continuously-differentiable function of $\varcomp\in\varcompparamspace$.
The dimension, $\numcomponents$, of $\varcomp$ is typically small.
The parametric bootstrap (\cref{sec:bootstrap}) will require minimizing
$\nrll(\varcomp)$ many times for different bootstrapped datasets and 
hence this minimization must be fast and stable. 
Bootstrapping will also require fitting the model based on \cref{eqn:varcompmodelrotated} with $\varcompconstr_1=\zero$
and hence cannot restrict $\varcompidx_j\geq0$. Instead, the optimization must account for the spectrahedral constraint that $\varcomp\in\varcompparamspace$.

\citet{bates2015fitting} recommend derivative-free box-constrained
minimization, pointing out that computation of derivatives in this
problem is difficult may result in a slower minimization.
However, in our present setup the constraint that $\varcomp\in\varcompparamspace$ cannot be expressed as a simple box constraint.
We find that two modifications to the usual Newton 
iteration yield an appropriate algorithm for minimizing $\nrll(\varcomp)$, and further that
computation of first and second derivatives using appropriate sparse matrix algebra via the identities (\ref{eqn:blockdetinv}) and (\ref{eqn:sigmainvidentities}) described in \cref{sec:decomposition} is sufficiently fast.
Accordingly, we use a modified  Newton's method with two analytic derivatives to minimize $\nrll(\varcomp)$ over $\varcomp\in\varcompparamspace$.

With chosen starting value $\varcomp^0$ (\cref{subsec:startingvalues}), tolerance $\epsilon$, 
and $t\in\Nats$, Newton's method computes the iterates:
\begin{equation}\label{eqn:basicnewton}
\varcomp^{t+1} = \varcomp^t - \left\{\nabla^2_{\varcomp\varcomp\Tr}\nrll(\varcomp^t)\right\}\inv\nabla_{\varcomp}\nrll(\varcomp^t),
\end{equation}
and the returned minimum is $\varcompmle = \varcomp^T$ for $T=\min\{t\in\Nats:\norm{\nabla_{\varcomp}\nrll(\varcomp^t)}_\infty < \epsilon\}$.

$\nrll(\varcomp)$ is not convex in $\varcomp$ for all $\nrs$.
At each iteration
the chosen search direction may not be a descent direction and
any mode found may be a local minimum, maximum, or saddle point.
Further, the constraint that $\varcomp\in\varcompparamspace$
is not enforced by \cref{eqn:basicnewton}.
We check the Hessian spectrum at convergence to ensure we have found a local minimum. 
While there is no theoretical guarantee that this is a global minimum, 
our simulations in \cref{sec:simulations} provide empirical evidence that these local minima at least have the desired statistical properties.
We modify the Newton iteration in two ways to address the remaining two challenges.

First, we compute the eigendecomposition of the Hessian,
$
\nabla^2_{\varcomp\varcomp\Tr}\nrll(\varcomp^t) = \mb{B}_t\mb{\Lambda}_t\mb{B}_t\Tr,
$
where $\mb{\Lambda}_t = \diag(\lambda_{1,t},\ldots,\lambda_{\numcomponents,t})$ holds the ordered eigenvalues $\lambda_{1,t}\geq,\ldots,\geq\lambda_{\numcomponents,t}$.
We replace $\nabla^2_{\varcomp\varcomp\Tr}\nrll(\varcomp^t) $ by $\mb{B}_t\mb{\Lambda}^{+}_t(\kappa)\mb{B}_t\Tr$ in each iteration of \cref{eqn:basicnewton}, where $\mb{\Lambda}^{+}_t(\kappa) = \text{diag}(\lvert\lambda_{1,t}\rvert+\kappa,\ldots,\lvert\lambda_{\numcomponents,t}\rvert+\kappa)$ for a small fixed $\kappa$.
Using absolute eigenvalues means that when $\lambda_{j,t}\ll0$ a step of the appropriate length is taken but the
direction is reversed from ascent to descent, and when $\lambda_{j,t}\approx0$ a fixed ``gradient step'' of large length $\approx\kappa\inv$ is taken in
the direction of the negative gradient.

Second, we find that a simple backtracking modification to \cref{eqn:basicnewton} is sufficient to account for the nontrivial constraint that $\varcomp\in\varcompparamspace$.
Observe that $\nrll(\varcomp)$ is a \emph{barrier} for $\varcompparamspaceU$ with $\lim_{\varcomp\to\partial\varcompparamspace}\nrll(\varcomp) = \infty$ where $\partial\varcompparamspace$
is the boundary of $\varcompparamspace$.
This means that as $\varcomp$ approaches $\partial\varcompparamspace$
the gradient of $\nrll(\varcomp)$ eventually changes sign and
$\varcomp$ switches directions away from $\partial\varcompparamspace$.
This in turn
implies that the only way that an iteration can escape into $\varcompparamspace^\complement$ is if it passes over a point where
the directional derivative changes sign, i.e. a point $\varcomp^{t+s} = \varcomp^t + s\left\{\nabla^2_{\varcomp\varcomp\Tr}\nrll(\varcomp^t)\right\}\inv\nabla_{\varcomp}\nrll(\varcomp^t)$ for $s\in[0,1]$ with $\langle{\grad\nrll(\varcomp^{t+s})},{\left\{\nabla^2_{\varcomp\varcomp\Tr}\nrll(\varcomp^t)\right\}\inv\nabla_{\varcomp}\nrll(\varcomp^t)}\rangle=0$.
Step-halving is therefore sufficient to account for the nontrivial constraint that $\varcomp\in\varcompparamspace$. While $\varcomp^t\in\varcompparamspace$
but $\varcomp^{t+1}\notin\varcompparamspace$, simply replace $\varcomp^{t+1}$ by $\varcomp^t + (\varcomp^{t+1} - \varcomp^t)/2$.

With an algorithm in hand for minimizing $\nrll(\varcomp)$
we compute 
\begin{equation}\label{eqn:varcompmle}
\varcompmle = \argmin_{\varcomp\in\varcompparamspace}\nrll(\varcomp) = \argmin_{\varcomp\in\Reals^d}\nrll(\varcomp)
\end{equation}
using the modified Newton iteration.
The second equality in \cref{eqn:varcompmle} follows from the constraint $\varcomp\in\varcompparamspace$ being built into $\nrll(\varcomp)$ through its
barrier property previously described, since $\nrll(\varcomp) = \infty$ for $\tau\in\varcompparamspace^\complement$.

\subsection{Starting values}\label{subsec:startingvalues}

Starting values are obtained via the method of moments by equating sequential sums of 
squares to their expectations under \cref{eqn:varcompmodel}
and then solving the resulting low-dimensional, triangular linear system for $\varcomp$.
For some cases such as balanced designs these starting values are exact and no modified Newton iterations are required. 
In cases where they are not exact, our simulations (\cref{sec:simulations}) and data analysis (\cref{sec:dataanalysis}) 
seem to indicate that these starting values tend to result in very few iterations being required for the modified Newton's method to converge.

For $j=1,\ldots,d$ let $\sumsquare{j}=\norm{\mb{U}_{j-1}\Tr\obsall}_2^2 - \norm{\mb{U}_{j}\Tr\obsall}_2^2$, where
$\mb{U}_j$ is an orthonormal basis for $\text{ker}[\mb{X}, \mb{V}_1:\cdots:\mb{V}_j]$.
Under the model \cref{eqn:varcompmodel} its expectation is
\begin{equation}\label{eqn:expectedmeansquare}
\begin{aligned}
\EE\sumsquare{j} &= \sigma^2\left[ \tr\left\{ \mb{U}_{j-1}\mb{U}_{j-1}\Tr\mb{\Sigma}(\varcomp)\right\} - \tr\left\{ \mb{U}_{j}\mb{U}_{j}\Tr\mb{\Sigma}(\varcomp)\right\}\right], \\
&= \sigma^2\left[ (r_{j-1} - r_j) + \sum_{k=j}^{d}\varcompidx_k\tr\left\{\mb{U}_{j-1}\mb{V}_k\mb{U}_{j-1}\right\} - \sum_{k=j+1}^{d}\varcompidx_k\tr\left\{\mb{U}_{j}\mb{V}_k\mb{U}_{j}\right\} \right],
\end{aligned}
\end{equation}
where we use \cref{eqn:covmat} and the fact that $\mb{U}_j\Tr\mb{V}_k = 0$ when $j\geq k$ by definition, and $r_j = \text{rank}(\mb{U}_j)$.
Setting $\sumsquare{j} = \EE\sumsquare{j}$ yields the $\numcomponents\times\numcomponents$ triangular
system of equations $\mb{s} = \mb{M}\varcomp^0$
where $s_j = \sumsquare{j}/\sigma^2 + r_j - r_{j-1}$
and
$$
M_{ij} = \begin{cases} 
0 & i < j, \\
\tr\left(\mb{U}_{j-1}\Tr\mb{V}_j\mb{U}_{j-1}\right) & i = j, \\
\tr\left(\mb{U}_{j-1}\Tr\mb{V}_i\mb{U}_{j-1}\right) - \tr\left(\mb{U}_{j}\Tr\mb{V}_i\mb{U}_{j}\right) & i > j.
\end{cases}
$$
Substituting $\widehat{\sigma}^2 = \norm{\mb{U}_d\Tr\obsall}_2^2 / (\numobs - p)$ for $\sigma^2$ and solving
for $\varcomp^0$ yields starting values for Newton's method.


\section{Parametric bootstrap hypothesis test}\label{sec:bootstrap}

Consider now a parametric bootstrap for testing $H_0:\lcmatrix\varcomp=\zero$ based on \cref{eqn:varcompmodelrotated}.
Fix $B\in\Nats$. 
Let $\mb{z}^{*}_1,\ldots,\mb{z}^{*}_B\sim\normaldist\left\{\zero, \mb{U}_X{}\Tr\mb{\Sigma}(\varcompmlenull)\mb{U}\right\}$ independently, and
$\nrsboot_b = \mb{z}^{*}_b / \norm{\mb{z}^{*}_b}_2$.
Let $\varcompmleboot_b$ be the maximum likelihod estimator based on $\nrsboot_b$, $b=1,\ldots,B$.
The parametric bootstrap $p$-value for testing $H_0:\lcmatrix\varcomp=\zero$ against the two-sided
alternative that $\lcmatrix\varcomp\neq\zero$ 
is $\pboot = \numboot\inv\sum_{b=1}^\numboot\mathbbm{1}\{\nrll(\varcompmleboot_b) \geq \nrll(\varcompmle)\}$.
The one-sided $p$-value for testing against the alterntive that $\lcmatrix\varcomp\in\mathcal{C}$ for a set $\mathcal{C}\subset\Reals^{\numcomponents-\numcomponents_0}$
is $\pboot = \numboot\inv\sum_{b=1}^\numboot\mathbbm{1}\{\nrll(\varcompmleboot_b) \geq \nrll(\varcompmle)\}\mathbbm{1}\{\varcompmleboot_b \in\mathcal{C}\}$.

The distribution of $\mb{z}_b$ under $H_0$ is not free of parameters. The variance components must be estimated under $H_0$.
To compute these estimates we re-use the modified Newton algorithm of \cref{sec:newton} and the rotated model from \cref{eqn:varcompmodelrotated}
to obtain
\begin{equation}\label{eqn:varcompmleconstr}
\varcompmlenull = \argmin_{\varcomp\in\varcompparamspace:\lcmatrix\varcomp=\zero}\nrll(\varcomp) = \mb{Q}_2\times\left[\argmin_{\varcompconstr_2\in\Reals^{d-r}}\nrll(\mb{Q}_2\varcompconstr_2) \right].
\end{equation}
However, the fact that $\varcompmlenull$ is estimated from the data means that the parametric bootstrap test is not guaranteed to
have the correct size. 
We discuss this further in the context of our empirical results in \cref{sec:simulations}.

Sampling from an $(\numobs-p)$-dimensional Gaussian distribution requires decomposition
of its $(\numobs-p)$-dimensional covariance matrix at cost $O(\numobs^3)$.
Even in models where $\mb{\Sigma}(\varcompmlenull)$ is sparse, $\mb{U}_X\Tr\mb{\Sigma}(\varcompmlenull)\mb{U}_X$
will be dense.
It is therefore desirable to avoid naively computing and decomposing
$\mb{U}_X\Tr\mb{\Sigma}(\varcompmlenull)\mb{U}_X$ when sampling the $\mb{z}_b$.
The decompositions in \cref{sec:decomposition} are reused to obtain an efficient algorithm for sampling $\mb{z}_1,\ldots,\mb{z}_B$.
A consequence of \cref{eqn:sigmadecomp} is that
if $\mb{\omega}\sim\normaldist(\zero,\mb{I}_{\numobs}) = (\mb{\omega}_1\Tr,\mb{\omega}_2\Tr)\Tr$ where $\text{dim}(\mb{\omega}_1) = \componentdim$,
then for any $\varcomp\in\varcompparamspace$,
\begin{equation}\label{eqn:adjustsample}
\mb{P}_Z\begin{pmatrix} \mb{L}(\varcomp)\mb{\omega}_1 \\ \mb{\omega}_2\end{pmatrix} \sim \normaldist\left\{\zero, \mb{\Sigma}(\varcomp)\right\},
\end{equation}
where $\mb{P}_Z$ and $\mb{L}(\varcomp)$ are given in \cref{eqn:sigmadecomp}. 
Since $\mb{P}_Z$ is constant for a given $\mb{Z}$ its Householder representation is computed once before computation of $\varcompmle$.
The Cholesky factor $\mb{L}(\varcompmlenull)$ is already available from the final iteration of
Newton's method for computing $\varcompmlenull$ (\cref{eqn:varcompmleconstr}). 
To generate $\nrsboot_b$ efficiently we sample $\mb{\omega}_b\sim\normaldist(\zero,\mb{I}_{\numobs})$,
adjust it via \cref{eqn:adjustsample} to obtain $\mb{\omega}^{*}_b = \mb{P}_Z((\mb{L}(\varcompmlenull)(\mb{\omega}_b)_{1:m})\Tr,(\mb{\omega}_b)_{(m+1):\numobs}\Tr)\Tr$, and then compute $\mb{z}_b^{*} = \mb{U}_X\Tr\mb{\omega}^{*}_b$. 
This procedure involves only triangular matrix multiplcation and Householder reflections.

\section{Empirical evaluation}\label{sec:simulations}

We investigate the size and power of the proposed bootstrap test
via simulations for nested and crossed designs.
Simulation studies for tests involving a single component would vary a single sample size and effect size.
The present simulations are more complicated as we vary multiple sample sizes and effect sizes
as well as design balance. 

Contrary to the single-component and simple hypothesis case, 
the null hypothesis is not free of unknown
parameters and hence data generation under the null requires fitting
the model. This is observed to lead to bootstrap tests with incorrect size in some highly unbalanced designs in which it is 
expected to be difficult to estimate variance components.

Computations were performed on the Digital Resource Alliance of Canada's Trillium cluster, whose hardware specifications are
available at \url{https://docs.alliancecan.ca/wiki/Trillium}.
Our \texttt{julia} package \texttt{varcomptest} was used and is available at \url{https://github.com/awstringer1/varcomptest-jl}. 
The code used to produce all results
is available at \url{https://github.com/awstringer1/varcomptest-paper-code/tree/main}.
Each of the nested and crossed simulation studies use $675$ parameter combinations
with $S=1000$ datasets per combination and $B=300$ parametric bootstrap samples per dataset for a total
of 202,500,000 runs of Newton's method per simulation study. No instances of non-convergence were observed.
The data analyses reported in \cref{sec:dataanalysis} and in \cref{tab:examples} run comfortably in seconds on a modern laptop.

Uniformity of simulated
bootstrap p-values $(\pboot)_1,\ldots,(\pboot)_S$ is used to evaluate the accuracy of the bootstrap approximation to the sampling 
distribution of the test statistic and is assessed by a KS test statistic computed as the maximum deviation between their empirical and theoretical cumulative distribution functions.
The Supplementary Materials include uniform QQ-plots and further discussion.
Empirical power is computed by simulating data at a grid of $\varcomp$
values and computing the proportion of the $S$ datasets at each chosen value of $\varcomp$ for which $H_0$ is rejected.
Results are communicated by drawing lines for each distinct value of $\tau_1$, with $\tau_1 - \tau_2$ or $\lvert\tau_1 - \tau_2\rvert$ on the $x$-axis.
All of the simulations considered test the equality of two components, with $\numcomponents_0=1$, $\numcomponents=2$ and $\lcmatrix = (1, -1)$ having dimension $1\times 2$.

\subsection{Nested random effects}\label{subsec:nested}

A three-level nested random effects model is
\begin{equation}\label{eqn:nestedmodel2}
\obs_{ijk} = \cov_{ijk}\Tr\regparam + u_i + v_{ij} + \epsilon_{ijk}; u_i \iidsim \normaldist(0,\sigma^2\tau_1), v_{ij}\iidsim\normaldist(0,\sigma^2\tau_2), \epsilon_{ijk}\iidsim\normaldist(0,\sigma^2),
\end{equation}
with $i=1,\ldots,m$ ``blocks'', $j=1,\ldots,n_i$ ``plots''
within block, and $k=1,\ldots,r_{ij}$ replications within
plot. The sample size is $N = r_{11} + \cdots + r_{mn_m}$.
When $r_{ij}\equiv r$ and $n_i\equiv n$ for all $i,j$
the design is referred to as balanced and $N = mnr$; otherwise, it is unbalanced.
We present results for an unbalanced design here;
results from a balanced design are given in the Supplementary Materials and are similar.
To achieve imbalance in the simulated datasets $m$ groups of size $n_i$ were sampled
where $n_i\sim\text{Unif}(2, 2n-2)$ independently so that
$\text{min}(n_i) = 2$ and $\mathbb{E}(n_i) = n$. 
Each subgroup consisted of $r_i$ replications
with $r_i\sim\text{Unif}(2, 2r-2)$, again so that
$\text{min}(r_i) = 2$ and $\mathbb{E}(r_i) = r$.
Further analysis in the Supplementary Materials shows KS statistics and uniform QQ plots of $p$-values
in which no clear pattern of deviation from the nominal size is observed.

\cref{fig:nestedpowertwoside} shows the empirical power against the two-sided alternative that $\tau_1 \neq \tau_2$.
Each line corresponds to one fixed value of $\tau_1$ and the $x$-axis corresponds to the absolute difference $\lvert\tau_1 - \tau_2\rvert$
for each simulated dataset. 
The test appears to have very little power with the smallest data sizes of $m=20$ groups with $n=2$ observations per group
and $r=2$ replications per observation, 
increasing when any of the samples sizes or the effect size is increased.
We have included a range of sample and effect sizes
such that cases of $\approx0\%$ to $\approx100\%$ power are observed.

\begin{figure}
\centering
\includegraphics[width=5in, height=7in]{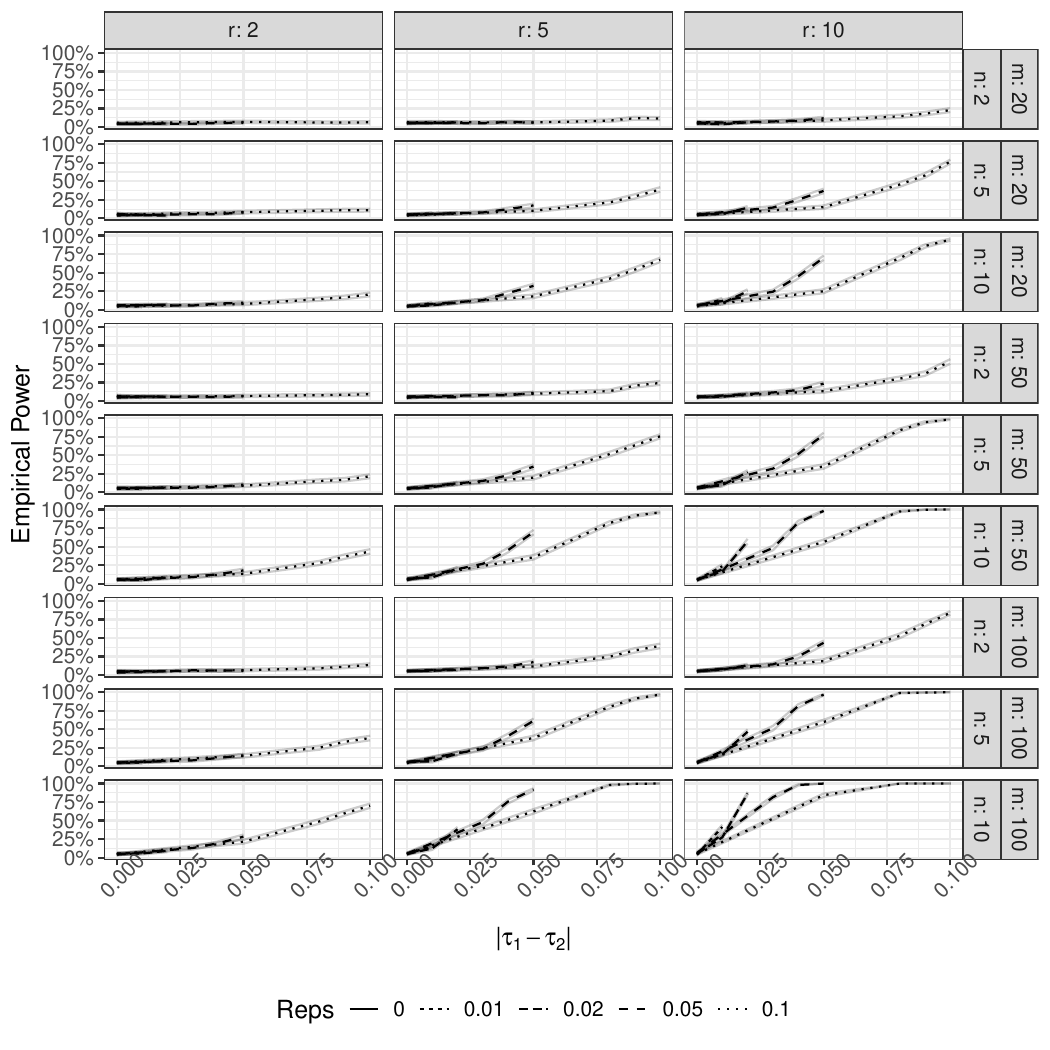}
\caption{Empirical power (with Monte Carlo standard error bands) for testing $H_0: \tau_1=\tau_2$ against the two-sided alternative $H_1:\tau_1\neq\tau_2$ in $1000$ simulated datasets from the nested model in \cref{eqn:nestedmodel2} for each choice of number of blocks $m$,
average block size $n$, average number of within-plot replications $r$, and variance components $\tau_1,\tau_2$ values in an unbalanced design. 
Power increases as the sample size increases through any of the 
number of blocks, average number of plots, or average number of replications, as well as as the effect size increases.}
\label{fig:nestedpowertwoside}
\end{figure}

\subsection{Crossed random effects}\label{subsec:crossed}

A crossed random effects model is
\begin{equation}\label{eqn:crossedmodel2}
\obs_{ijk} = \cov_{ijk}\Tr\regparam + u_i + v_{j} + \epsilon_{ijk}; u_i \iidsim \normaldist(0,\sigma^2\tau_1), v_{j}\iidsim\normaldist(0,\sigma^2\tau_2), \epsilon_{ijk}\iidsim\normaldist(0,\sigma^2),
\end{equation}
where $i \in \{1,\ldots,m\}$ and $j \in \{1,\ldots,n\}$ and $k\in\{1,\ldots,r_{ij}\}$.
The sample size is $N = r_{11} + \cdots + r_{mn_m}$.
Each distinct pair $(i,j)$ may or may not be present in the data.
When all such pairs are present, $m=n$, and $r_{ij}\equiv r$ is constant, the design is referred to as balanced; otherwise it is unbalanced.
We consider a balanced design with sample size $N=mn$ and each $(i,j)$
pair occurring $r=1$ times.
We consider an unbalanced design with the strength of the imbalance controlled by
a parameter $0\leq \rho<1$. The unbalanced design is created by sampling $N$ values 
from a $\text{N}(0, \mb{\Omega}(\rho))$ distribution where $\mb{\Omega}(\rho)$ is a $2\times 2$ correlation matrix with off-diagonal equal to $\rho$, and
then obtaining index pairs $(i,j)$ by transforming the univariate
marginals of this multivariate Normal to discrete 
uniforms on $1,\ldots,m$ and $1,\ldots,n$. 
The average sample size is $N=mn$ but $r_{ij}$ is random and may be $0$ for certain $(i,j)$ pairs.

\begin{figure}
\centering
\includegraphics[width=5in, height=7in]{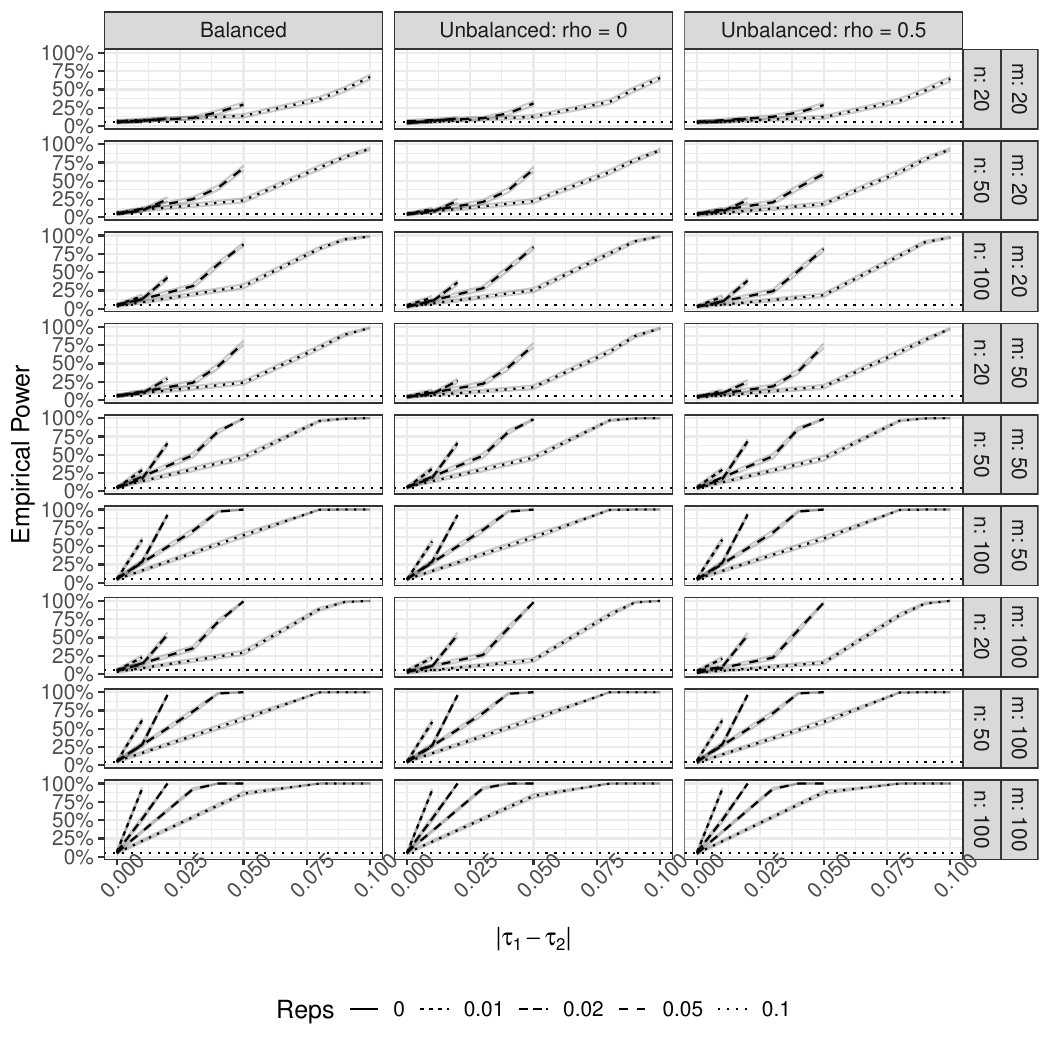}
\caption{Empirical power for testing $H_0: \tau_1=\tau_2$ against the two-sided alternative $H_1:\tau_1\neq\tau_2$ in $1000$ simulated datasets from the crossed model in \cref{eqn:crossedmodel2} for each choice of the two sample sizes $m$ and $n$ and common $\tau$ values, and for a balanced design as well as the unbalanced design with $\rho=0, 0.5$. 
In the designs with more unbalanced sample sizes the power decreases for less balanced effects.
}
\label{fig:crossedpowertwoside}
\end{figure}

We simulate $S=1000$ datasets from \cref{eqn:crossedmodel2} at different $m$, $n$, and $\rho$ and 
obtain the empirical power shown in \cref{fig:crossedpowertwoside}.
The power does not appear to be materially affected by the design balance.
However, the Supplementary Materials show some evidence of the test deviating from nominal size for the more unbalanced designs. The two-sided test appears conservative in some cases, but the one-sided test is sometimes conservative and sometimes optimistic.

\section{Data analysis}\label{sec:dataanalysis}

\subsection{Overview of results}

We explore three examples from \cref{tab:examples} in \cref{subsec:motivation} in greater
detail. The parametric bootstrap reveals substantial variability
in the distribution of the maximum likelihood estimator under the
null hypothesis for two examples. This explains the failure to reject even in the presence of a large point estimate as seen in \cref{tab:examples} and underscores the need for a formal test of this hypothesis.
The final example shows a case where the hypothesis is 
rejected at this level.

\subsection{Chemical paste data}\label{subsection:paste}

A nested design reported by \citet{davies1947statistical} on 
the quality of two replicates of ten batches of a chemical 
paste product in three casks per batch for a sample size of $\numobs=60=2\times10\times3$ has the null hypotheses of zero batch and 
cask-within-batch variance components rejected by an $F$-test.
The point estimates of the variance components are $\widehat{\tau}_1 = 2.4$ and $\widehat{\tau}_2 = 12.4$ for a very large 
point estimated difference of $12.4 - 2.4 = 10$ and ratio of $12.4/2.4 = 5.17$.
However, the bootstrap p-value for the equality hypothesis of $0.17\pm0.024$
against the two-sided alternative and $0.13\pm0.021$ against the one-sided alternative that cask-within-batch exceeds batch variability, failing to reject at the $5\%$ level.
Despite this very large apparent discrepency between the two variance components,
there is insufficient power to conclude that the variability due to the two components are significantly different from each other. 

\begin{figure}
\centering
\begin{subfigure}[t]{0.48\textwidth}
	\includegraphics[width=3in, height=3in]{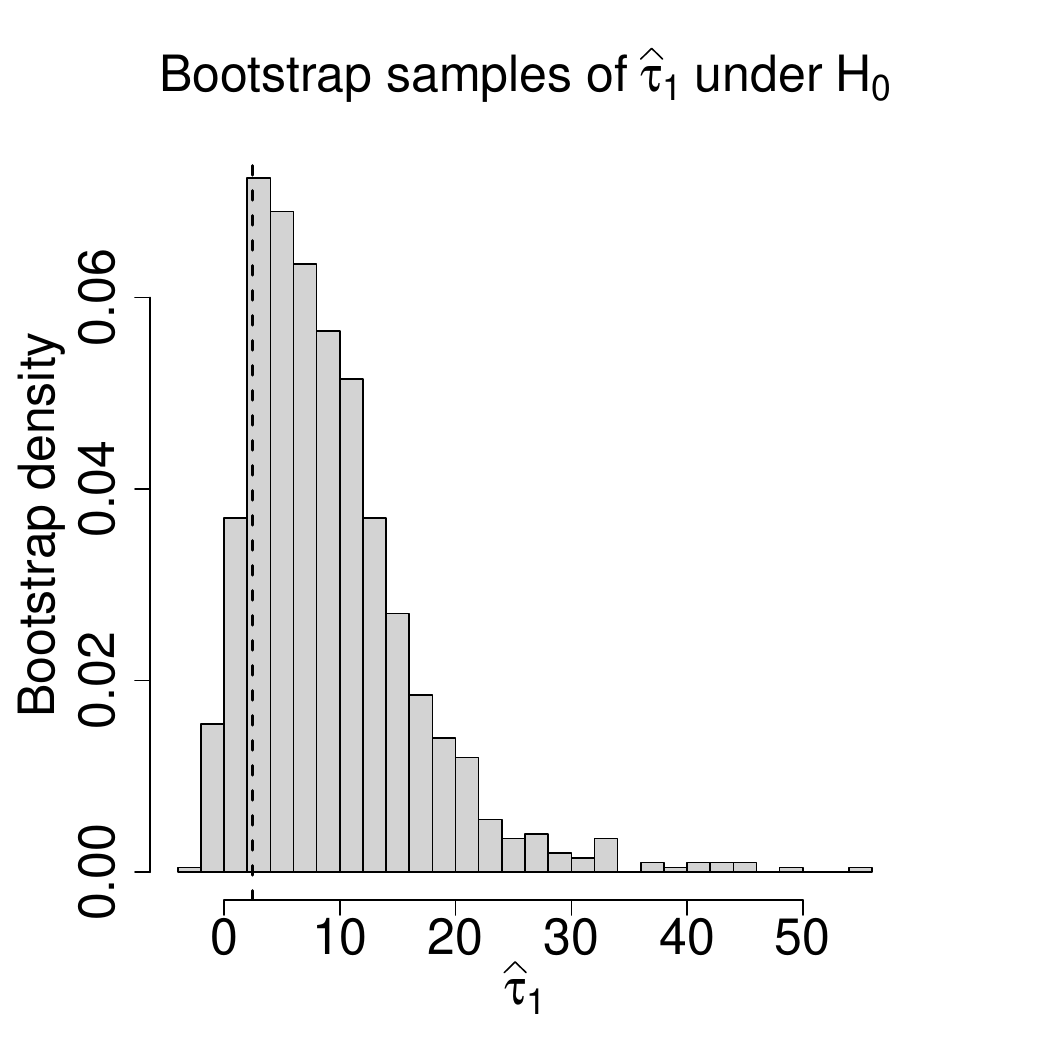}
	\caption{$\widehat{\tau}_1$}
	\label{subfig:pastetau1}
\end{subfigure}
\begin{subfigure}[t]{0.48\textwidth}
	\includegraphics[width=3in, height=3in]{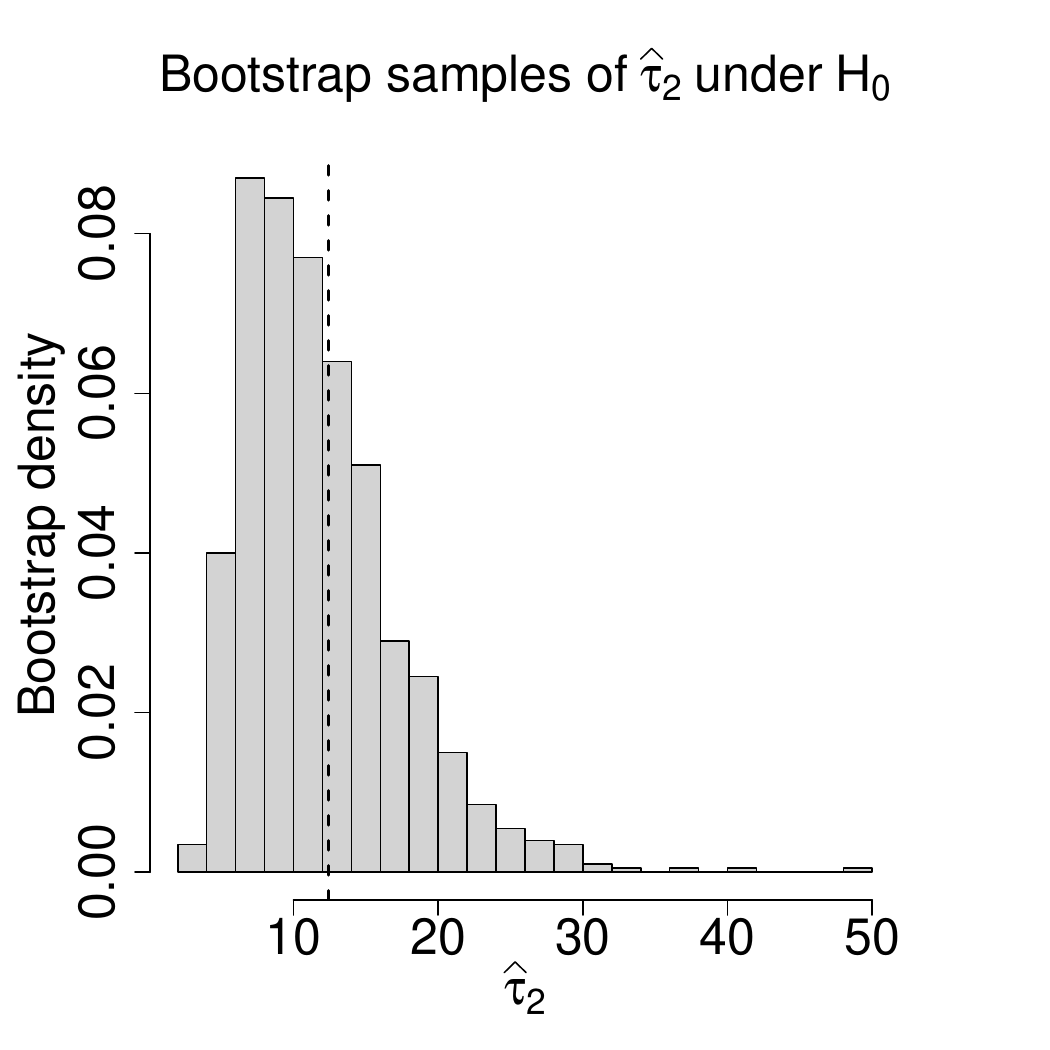}
	\caption{$\widehat{\tau}_2$}
	\label{subfig:pastetau2}
\end{subfigure}
\\
\begin{subfigure}[t]{0.48\textwidth}
	\includegraphics[width=3in, height=3in]{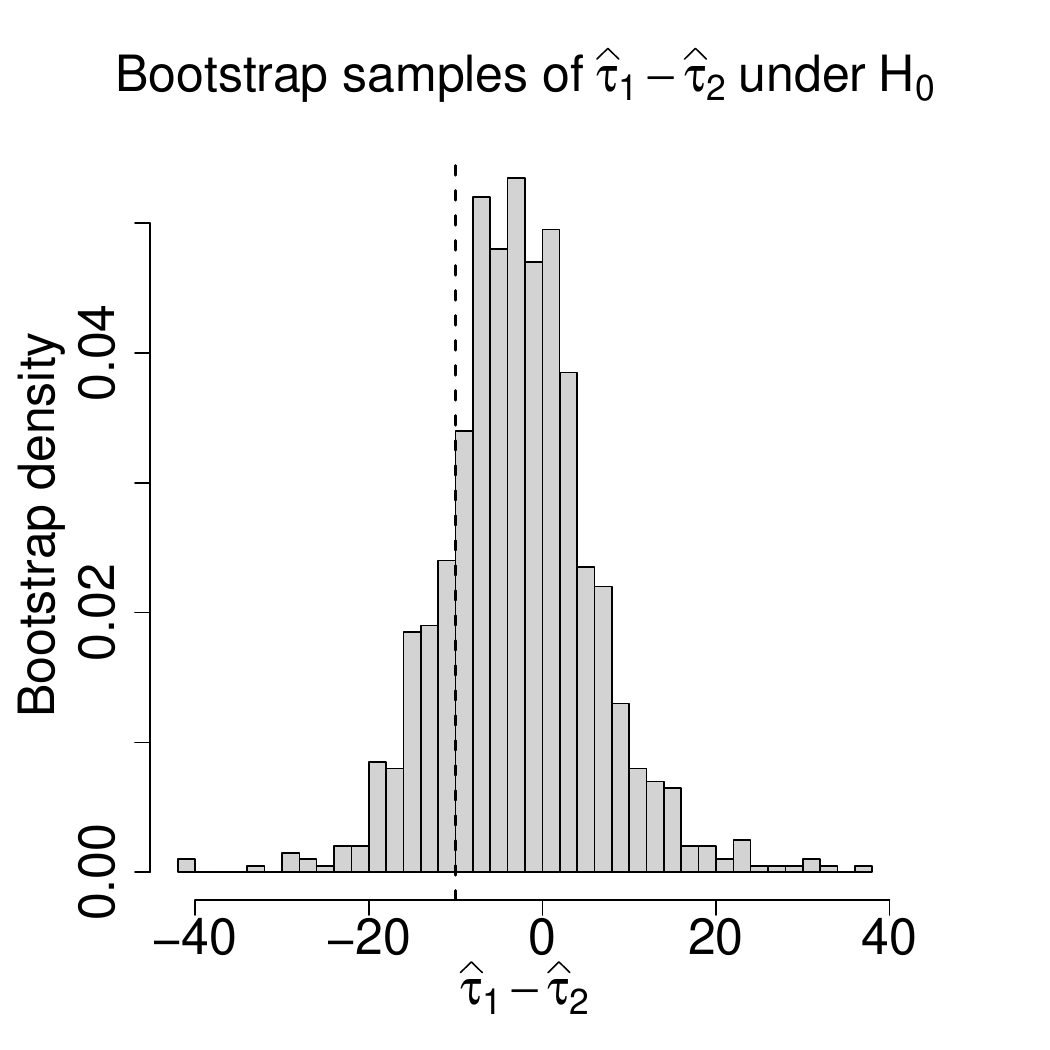}
	\caption{$\widehat{\tau}_1 - \widehat{\tau}_2$}
	\label{subfig:pastetaudiff}
\end{subfigure}
\begin{subfigure}[t]{0.48\textwidth}
	\includegraphics[width=3in, height=3in]{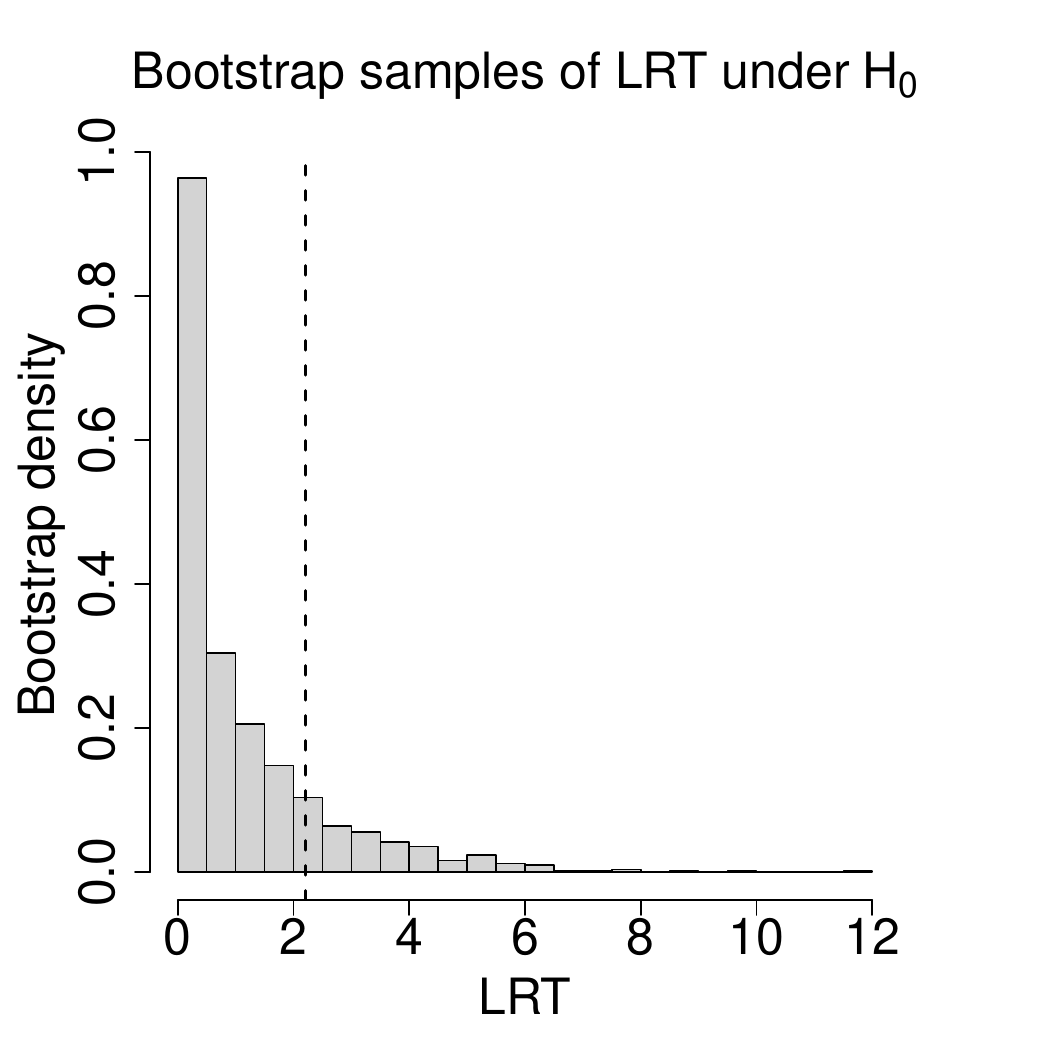}
	\caption{$\nrll(\varcompmle)$}
	\label{subfig:pastelrt}
\end{subfigure}
\caption{Bootstrapped sampling distributions of four functions of $\widehat{\varcomp}$ under $H_0:\tau_1=\tau_2$ for the chemical paste data of \cref{subsection:paste}.
The vertical dashed lines indicate the point estimates of each statistic. Substantial variability in these sampling distributions leads to a lack
of power for concluding that the apparently large estimated difference of $\widehat{\tau}_1 - \widehat{\tau}_2 = 10$ is statistically significant.
}
\label{fig:pasteshistograms}
\end{figure}

A benefit of using the bootstrap is the ability to more thoroughly investigate the estimated sampling 
distribution of the estimated variance components and functions of them under $H_0$.
\cref{fig:pasteshistograms} shows the bootstrap distributions of $\widehat{\tau}_1$, $\widehat{\tau}_2$, $\widehat{\tau}_1 - \widehat{\tau}_2$, and
$\nrll(\varcompmle)$.
We observe substantial variability and skew in the parameter distributions which is consistent with 
the lack of statistical significance of such a practically large estimated discrepancy between variance components.

\subsection{Semiconductor oxide thickness data}\label{subsection:oxide}

A further interesting example of a nested design is the data
reported by \citet{pinheiro2000mixed} on oxide thickness in
semi-conductor wafers. 
Three silicon wafers selected from eight lots (blocks) had oxide thickness measured at three sites each, for site (subplot) nested
within wafer (whole plot).
The sample size is $\numobs=72=3\times8\times3$. 
They state that in the original study the ``objective is to estimate the variance components to determine assignable causes of observed variability'' which strongly suggests an interest in
comparing variance components despite no test being available for this purpose.
\citet{pinheiro2000mixed} proceed to directly and explicitly compare variance components in their description of the data, specifically their Figure 4.4.
Despite estimated variance components of $10.34$ for lot and
$2.85$ for wafer-within-lot, a two-sided test of equality of
these components fails to reject at the $5\%$ level with a p-value of $0.12\pm0.021$. However, the p-value for the one-sided alternative that lot variation exceeds wafer-within-lot variation is $0.02\pm0.009$.
The language in the study description does not seem to suggest
a-priori interest in the one-sided hypothesis.

\begin{figure}
\centering
\begin{subfigure}[t]{0.48\textwidth}
	\includegraphics[width=3in, height=3in]{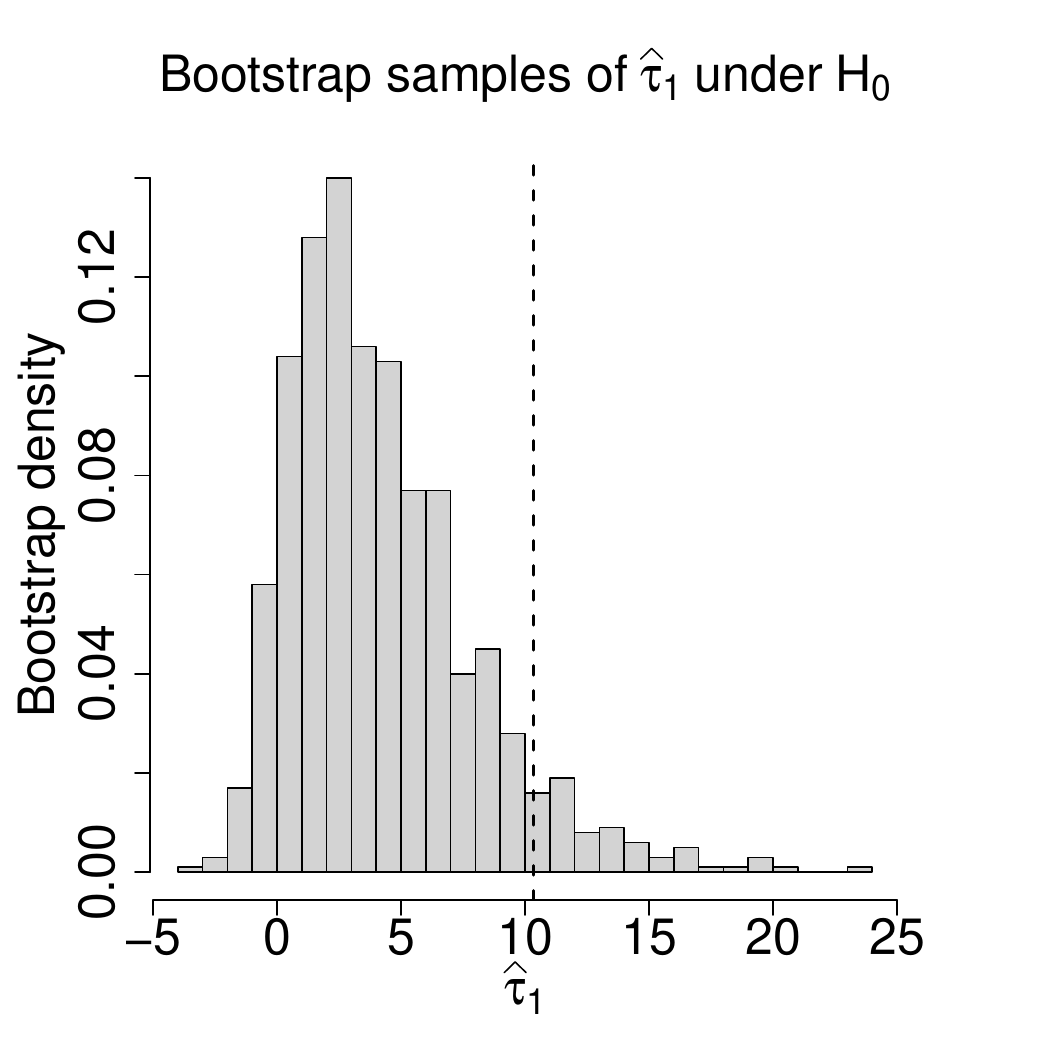}
	\caption{$\widehat{\tau}_1$}
	\label{subfig:oxidetau1}
\end{subfigure}
\begin{subfigure}[t]{0.48\textwidth}
	\includegraphics[width=3in, height=3in]{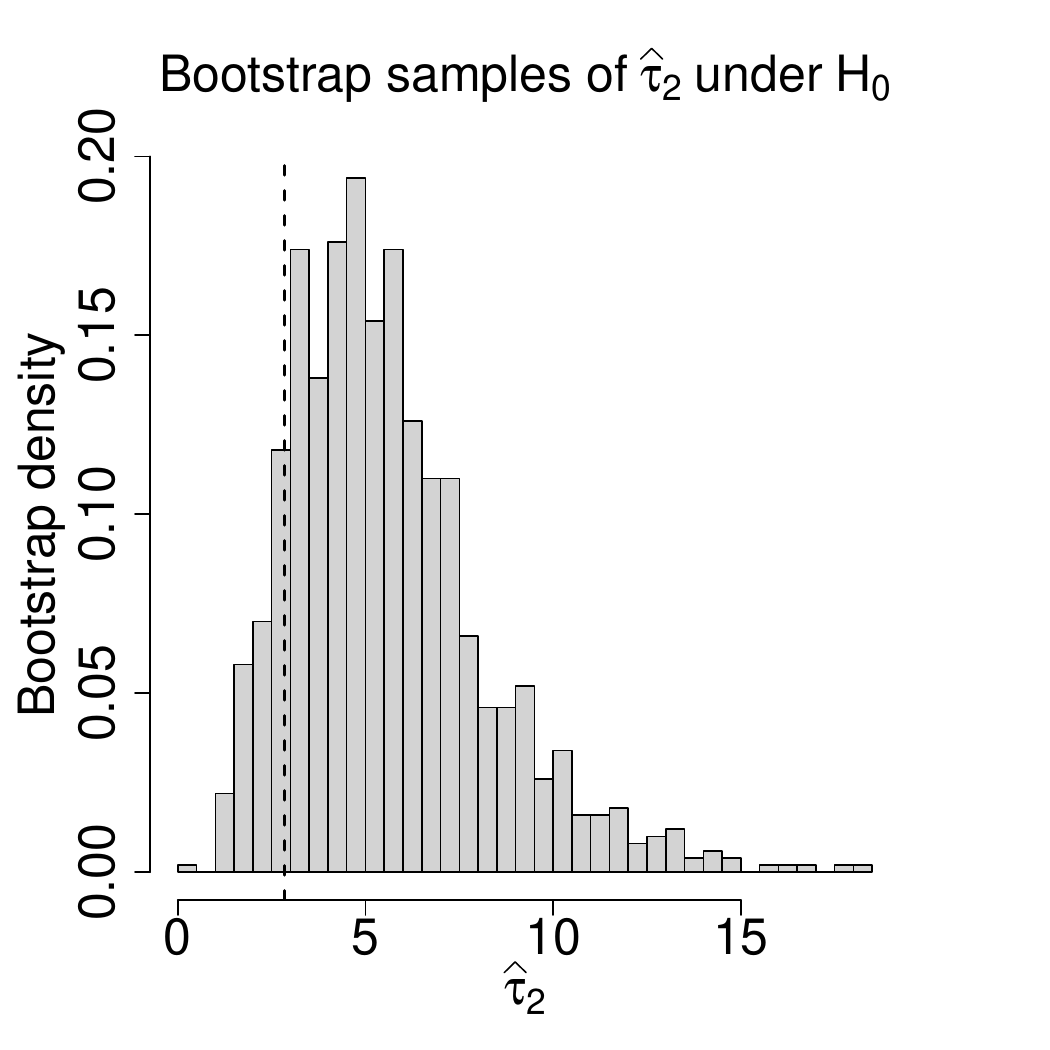}
	\caption{$\widehat{\tau}_2$}
	\label{subfig:oxidetau2}
\end{subfigure}
\\
\begin{subfigure}[t]{0.48\textwidth}
	\includegraphics[width=3in, height=3in]{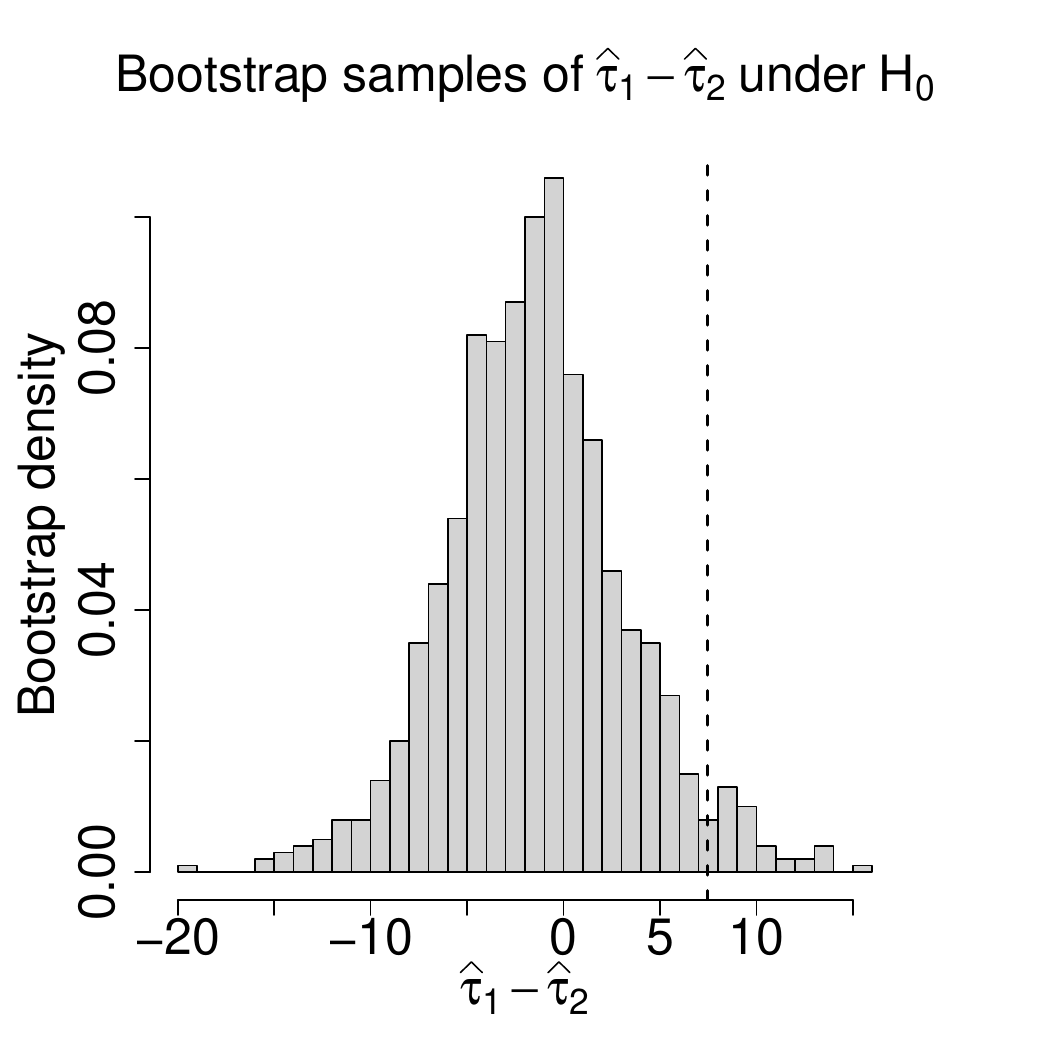}
	\caption{$\widehat{\tau}_1 - \widehat{\tau}_2$}
	\label{subfig:oxidetaudiff}
\end{subfigure}
\begin{subfigure}[t]{0.48\textwidth}
	\includegraphics[width=3in, height=3in]{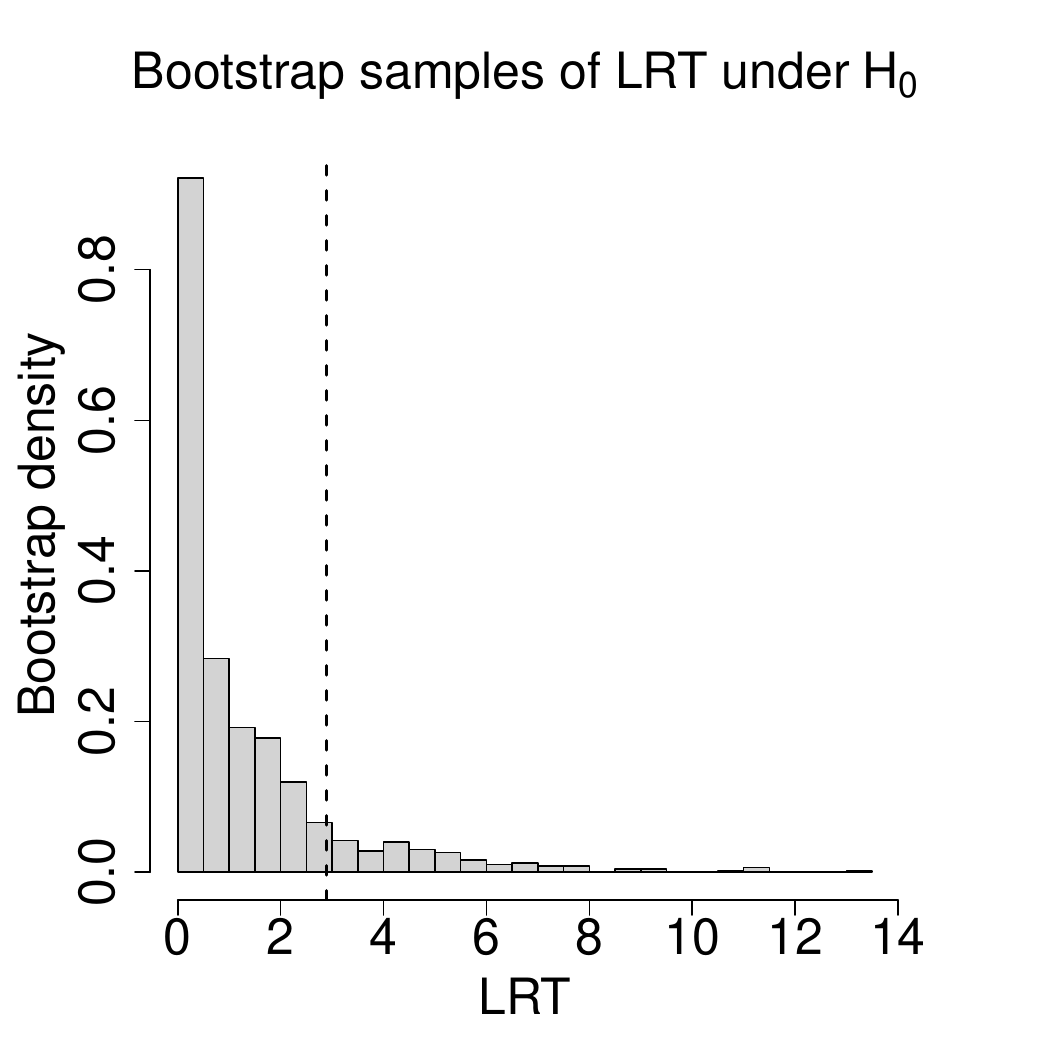}
	\caption{$\nrll(\varcompmle)$}
	\label{subfig:oxidelrt}
\end{subfigure}
\caption{Bootstrapped sampling distributions of four functions of $\widehat{\varcomp}$ under $H_0:\tau_1=\tau_2$ for the semiconductor oxide thickness data of \cref{subsection:oxide}.
The vertical dashed lines indicate the point estimates of each statistic. Substantial variability in these sampling distributions leads to a lack
of power, and the apparently large estimated difference of $\widehat{\tau}_1 - \widehat{\tau}_2 = 7.48$ is not statistically significant.
}
\label{fig:oxideshistograms}
\end{figure}

\cref{fig:oxideshistograms} shows the bootstrap distributions of $\widehat{\tau}_1$, $\widehat{\tau}_2$, $\widehat{\tau}_1 - \widehat{\tau}_2$, and
$\nrll(\varcompmle)$ under $H_0$.
We again observe substantial variability and skew in the parameter distributions which is consistent with 
the lack of statistical significance of such a practically large estimated discrepancy between variance components.

\subsection{Penicillin data}\label{subsection:penicillin}

\begin{figure}
\centering
\begin{subfigure}[t]{0.48\textwidth}
	\includegraphics[width=3in, height=3in]{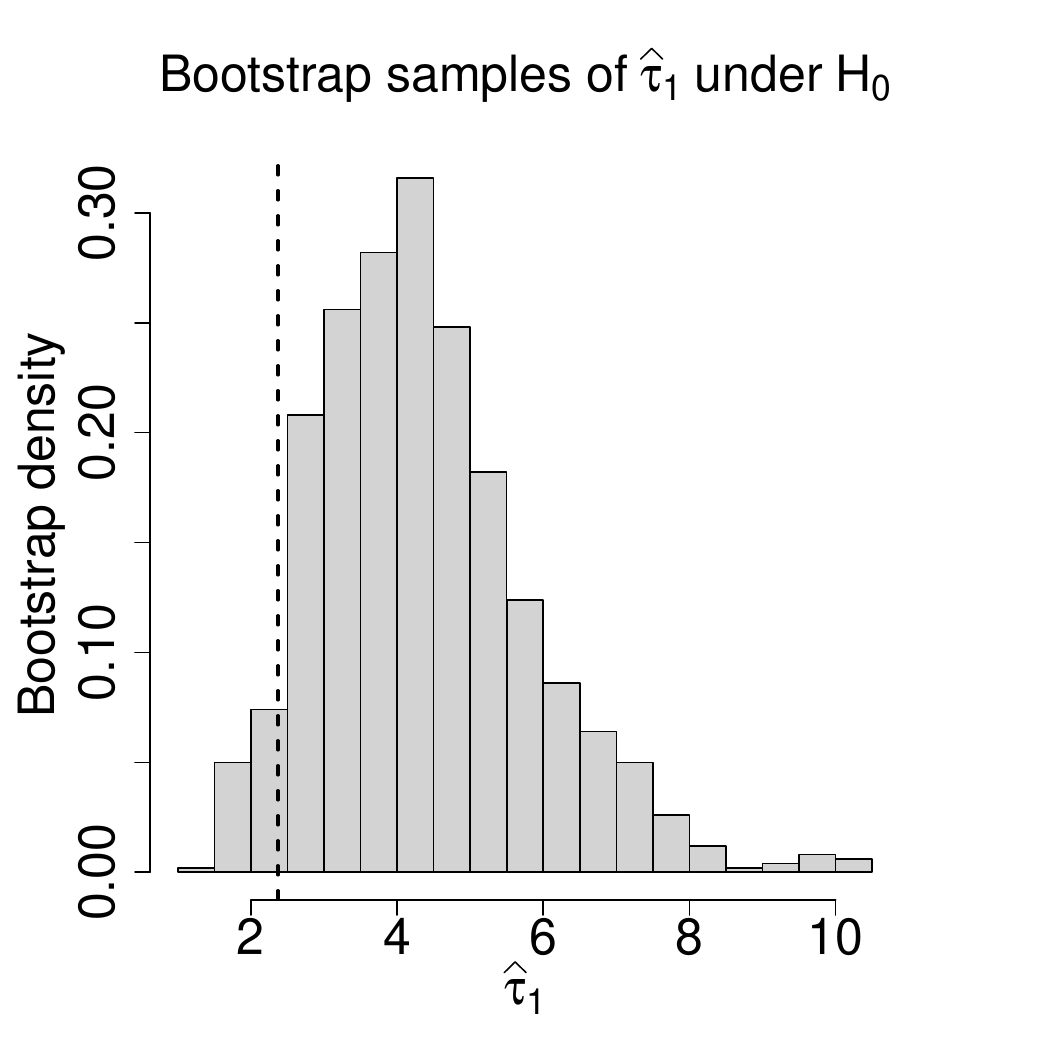}
	\caption{$\widehat{\tau}_1$}
	\label{subfig:oxidetau1}
\end{subfigure}
\begin{subfigure}[t]{0.48\textwidth}
	\includegraphics[width=3in, height=3in]{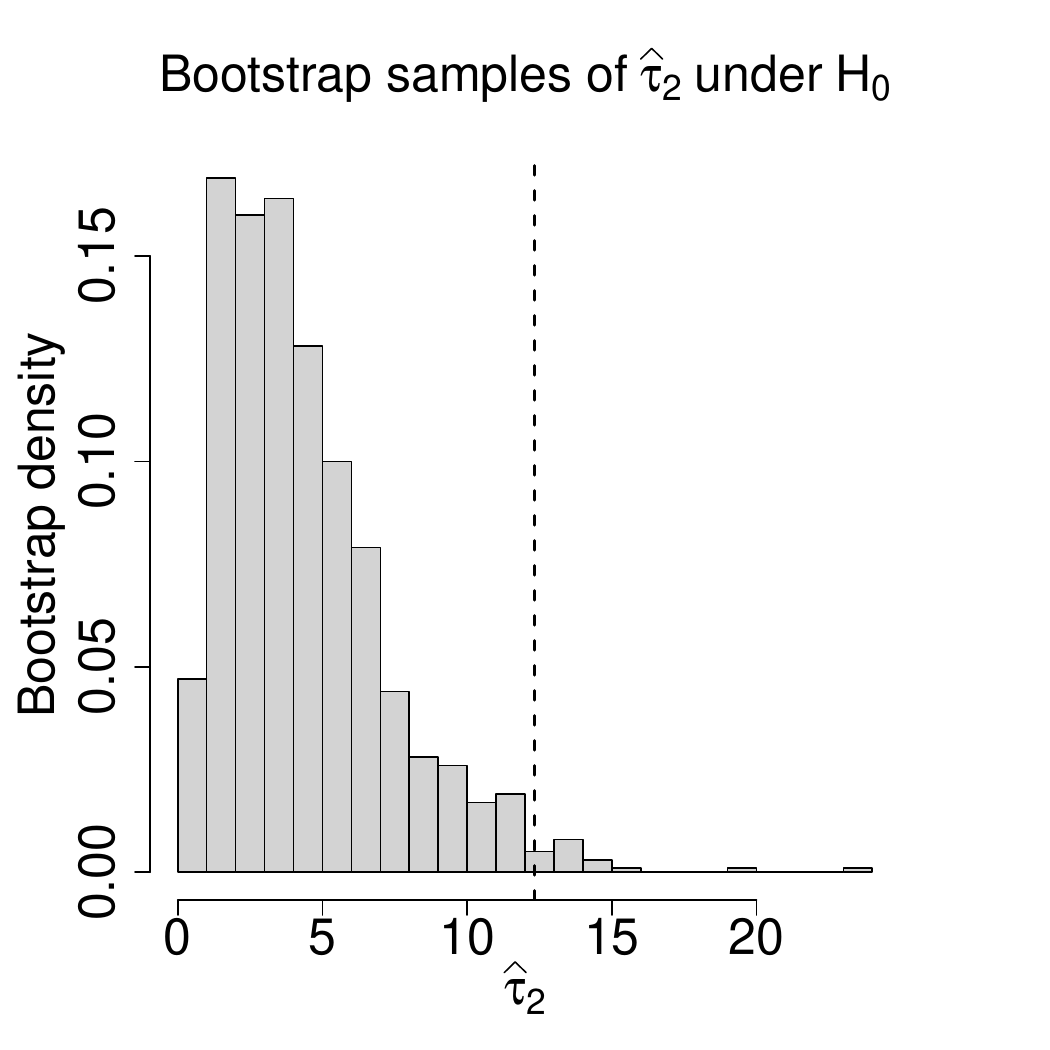}
	\caption{$\widehat{\tau}_2$}
	\label{subfig:oxidetau2}
\end{subfigure}
\\
\begin{subfigure}[t]{0.48\textwidth}
	\includegraphics[width=3in, height=3in]{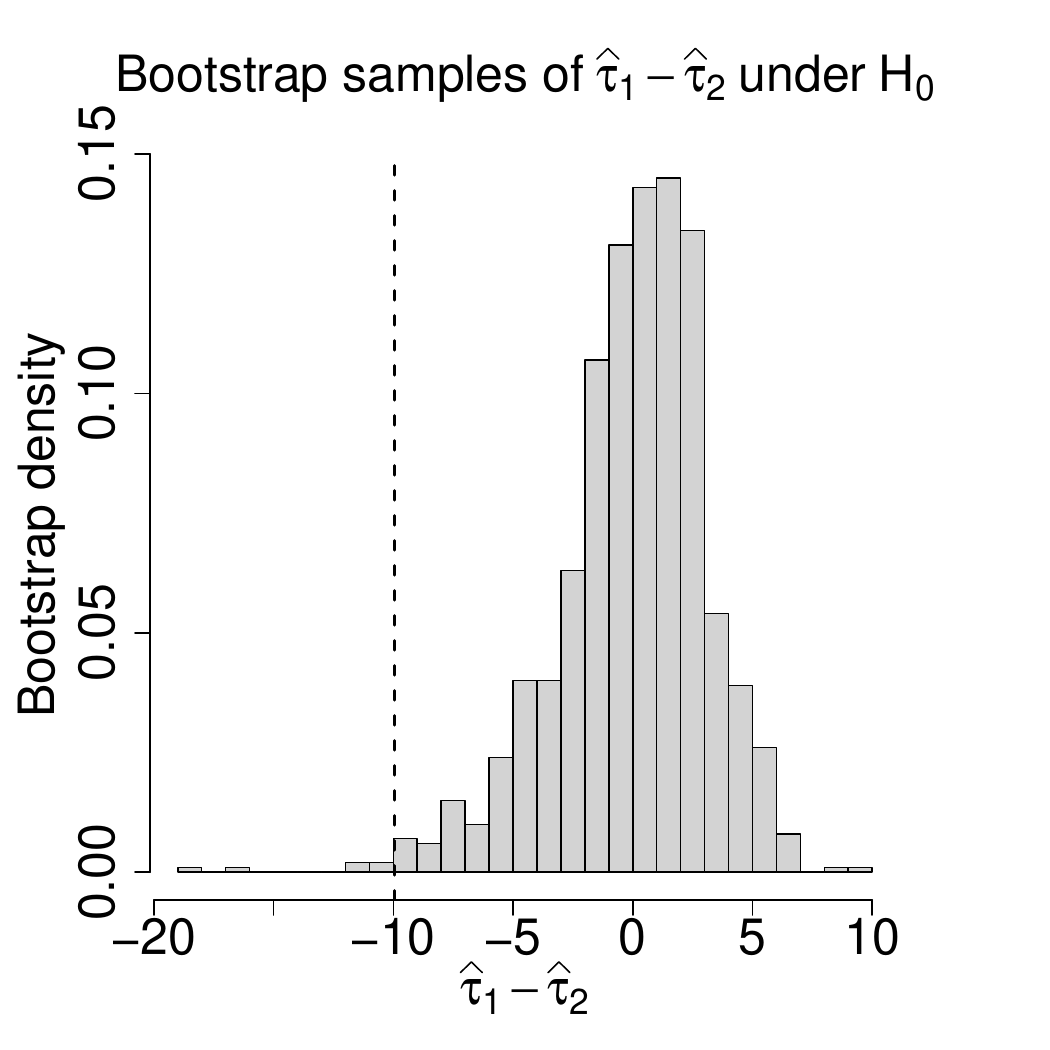}
	\caption{$\widehat{\tau}_1 - \widehat{\tau}_2$}
	\label{subfig:oxidetaudiff}
\end{subfigure}
\begin{subfigure}[t]{0.48\textwidth}
	\includegraphics[width=3in, height=3in]{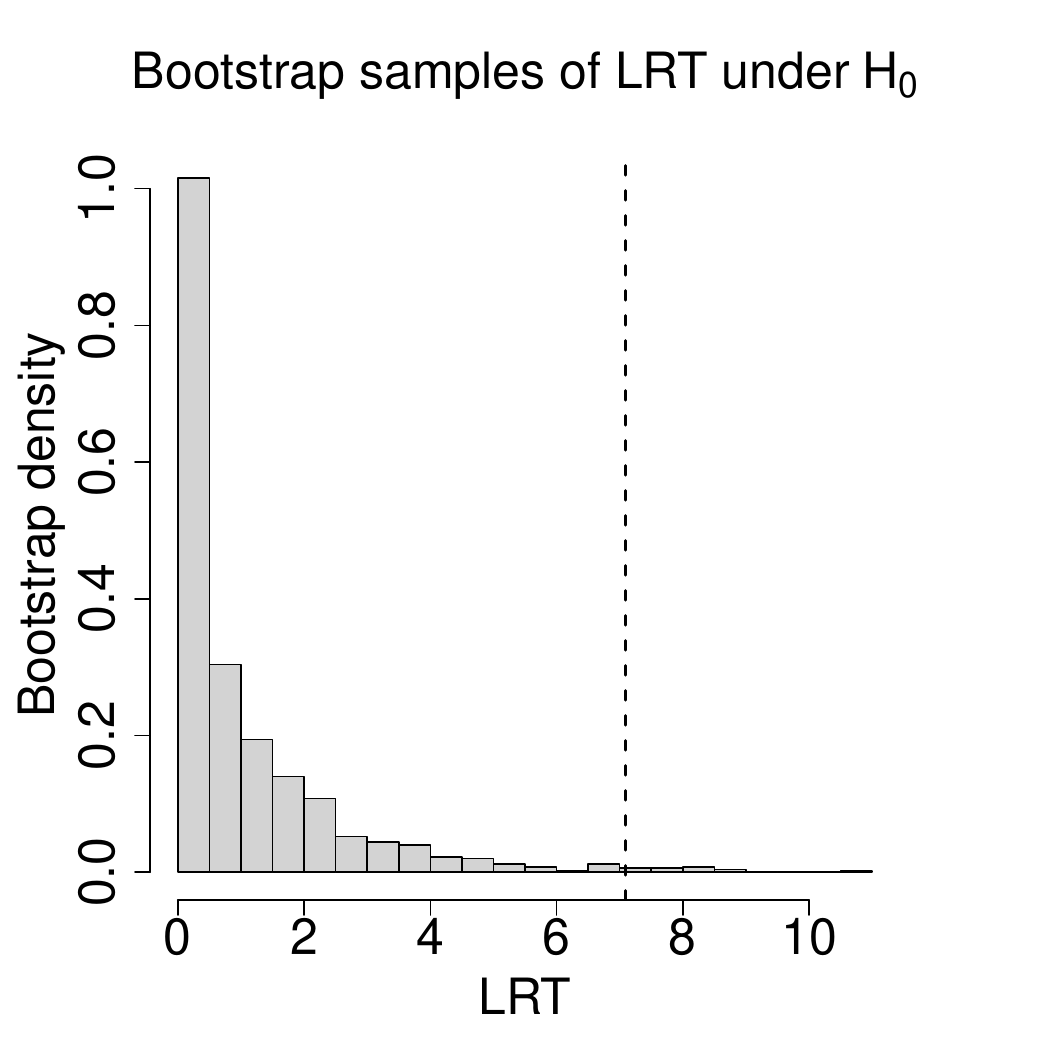}
	\caption{$\nrll(\varcompmle)$}
	\label{subfig:oxidelrt}
\end{subfigure}
\caption{Bootstrapped sampling distributions of four functions of $\widehat{\varcomp}$ under $H_0:\tau_1=\tau_2$ for the penicillin data of \cref{subsection:penicillin}.
The vertical dashed lines indicate the point estimates of each statistic. 
Only moderate variability in these sampling distributions leads to enough
power to conclude that the large estimated difference of $\widehat{\tau}_1 - \widehat{\tau}_2 = -9.97$ is statistically significant.
}
\label{fig:penicillinshistograms}
\end{figure}

An example where the test of equality of two variance components with a large point estimated difference is rejected is the 
fully crossed experiment on penicillin growth reported by \cite[Section 6.6]{
davies1972statistical}.
Six samples of penicillin solution were applied to 
each of $24$ plates for a total of $\numobs=144$ samples yields estimated variance components of $2.37$ for plate
and $12.34$ for sample for a large estimated difference of $2.37 - 12.34 = -9.97$.
The previous examples show that such an
observed difference cannot be trusted without a corresponding estimate of uncertainty.
However, in this example we find that our
new bootstrap test comfortably rejects the hypothesis of equality
against both the two- and one-sided alternatives.
\cref{fig:penicillinshistograms} shows that the bootstrap distributions are much less variable in this example than
was observed in \cref{fig:pasteshistograms,fig:oxideshistograms}.
The observed difference $\widehat{\tau}_1 - \widehat{\tau}_2$ is far enough in the tail of the boostrapped distribution of
$\tau_1 - \tau_2$ that it gives strong evidence against the null hypothesis.

\renewcommand{\appendixname}{Supplementary materials}
\appendix

\section{Further details on motivating examples}

This section contains text quoted from each reference in Section 1.3 where the authors describe their desire to test hypotheses involving
linear combinations of variance components.

\subsection{\citet{nye2004large}}

This work examines several variance component models for students success in reading and math. The variance components correspond to the factors including: ``school'', ``teacher'', ``classroom'', and ``student'', sometimes interacted with covariates such as socioeconomic status or grade. The exact form of each model fit varies across the different analyses in the paper. In many cases, the authors' statements about their results correspond to testing linear combinations of variance components. For example,
\citet{nye2004large} state:
\begin{center}
\begin{minipage}{0.8\textwidth}
\textit{\enquote{We found there was more variance in teacher effects in low-SES [socioeconomic status] schools than high-SES schools.}}
\end{minipage}
\end{center}
\begin{center}
\begin{minipage}{0.8\textwidth}
\textit{\enquote{The variation due to differences among teachers is substantial in comparison to the variation between schools.}}
\end{minipage}
\end{center}
These both correspond to alternative hypotheses for variance components of the form $H_A: \tau_1>\tau_2$, in the first example where the $\tau$ are the variance components for teacher effects within differing socioeconomic statuses, and in the second example where the $\tau$ are the teacher and school variance components. For another model, they also report that:
\begin{center}
\begin{minipage}{0.8\textwidth}
\textit{\enquote{In reading, the between teacher variance component is over twice as large as between-school variance component at grade 2, and over three times as large at grade 3.}}
\end{minipage}
\end{center}
The authors' statements correspond to the null hypothesis $\lcmatrix\varcomp=\zero$ where
$$
\lcmatrix = \begin{pmatrix} 1 & -2 & 0 & 0 \\ 0 & 0 & 1 & -3 \end{pmatrix}.
$$
The implied alternative hypothesis is that $\lcmatrix\varcomp\in\mathcal{C}$ where
$$
\mathcal{C} = \left\{\varcomp\in\Reals^4: \varcompidx_1 > 2\varcompidx_2 \land \varcompidx_3 > 3\varcompidx_4 \right\}
$$
In this case the variance components are the teacher and school variance components within grade 2 and the teacher and school variance components within grade 3.

\subsection{\citet{messier2010traits}}
The variance components are $\varcomp = (\varcompidx_1,\varcompidx_2,\varcompidx_3,\varcompidx_4, \varcompidx_5)\Tr$.
They correspond to the factors: ``site'', ``plot'', ``species'', ``tree'', ``strata'' and ``leaf''. 
\citet{messier2010traits} state:
\begin{center}
\begin{minipage}{0.8\textwidth}
\textit{\enquote{The plot level shows virtually no variance despite high species turnover among plots and the size of within-species variation (leaf + strata + tree) is comparable with that of species level variation.}}
\end{minipage}
\end{center}
This corresponds to a null hypothesis of the form $H_0: (\tau_1 = 0 \land \tau_2+\tau_3+\tau_4=\tau_5)$.
The authors' statements correspond to the null hypothesis $\lcmatrix\varcomp=\zero$ where
$$
\lcmatrix = \begin{pmatrix} 1 & 0 & 0 & 0 & 0 \\ 0 & 1 & 1 & 1 & -1 \end{pmatrix}.
$$
The implied alternative hypothesis is that $\lcmatrix\varcomp\in\mathcal{C}$ where
$$
\mathcal{C} = \left\{\varcomp\in\Reals^5: \varcompidx_1 \neq  0 \lor \varcompidx_2 + \varcompidx_3 + \varcompidx_4 \neq \varcompidx_5 \right\}
$$

\subsection{\citet{hill2008data}}
A third example from genetics, \citet{hill2008data} compare the variability of complex traits, and how it can be partitioned into three sources of variability (additive, dominance, and epistatic). A comparison they include in their abstract is:
\begin{center}
\begin{minipage}{0.8\textwidth}
\textit{\enquote{We evaluate the evidence from empirical studies of genetic variance components and find that additive variance typically accounts for over half, and often close to 100\%, of the total genetic variance.}}
\end{minipage}
\end{center} 
The ``over half'' part of that statement corresponds to an alternative hypothesis of the form $H_A:(\tau_1>\tau_2+\tau_3)$.
The authors' statements correspond to the null hypothesis $\lcmatrix\varcomp=\zero$ where
$$
\lcmatrix = \begin{pmatrix} 1 & -1 & -1 \end{pmatrix}.
$$
The implied alternative hypothesis is that $\lcmatrix\varcomp\in\mathcal{C}$ where
$$
\mathcal{C} = \left\{\varcomp\in\Reals^3: \varcompidx_1 > \varcompidx_2 + \varcompidx_3 \right\}
$$

\subsection{Statistics literature examples}\label{subsec:examples}

Table 1 in the main text shows results from several examples in the classical literature on variance
component models in which a nested or crossed design with two
variance components rejects the hypothesis that each are zero
based on an F-test, but no test of equality of the components
to each other was previously available. We provide background and citations for those examples here.
All of the datasets are routinely available in \texttt{R} \citep{Rlanguage}.

A famous example of a blocked split plot design is the oat yields
data originally reported by \citet{yates1935complex} and more recently
analyzed by \citet[Ch. 10]{venables2002random}.
Four treatments (varieties of nitrogen fertilizer) were applied
to three plots of different varieties of oats in six different blocks (batches) for a total of $72 = 4\times 3\times 6$ yield measurements. An $F$-test for the batch and variety-within-batch effects rejects at the $5\%$ level the hypotheses that the block variation and the plot-within-block variation is zero.
Applying our new test for equality of these components gives a 
bootstrap p-value of $0.57\pm0.03$ against the two-sided 
alternative and $0.19\pm0.025$ against the one-sided 
alternative that block variation exceeds plot variation, using
$B = 1000$ bootstrap samples.
The alfalfa yield
data reported by \citet[Appendix A.1]{pinheiro2000mixed} show
essentially the same design and results as the oats data, rejecting
the hypothesis that both block and plot variability are zero
while failing to reject the equality hypothesis against both the two- and one-sided alternatives.

The classical literature also contains several examples of crossed designs for which the new test is relevant.
A study reported by \citet[Section 1.3, Appendix A.14]{pinheiro2000mixed}
on the efficiency of an industrial process measured productivity
by six workers each operating three machines three times in a fully crossed design with three replications with total sample size $3\times6\times3=54$ gives variance component estimates 
of $4.82$ for machine and $2.65$ for worker, both significantly
nonzero by an $F$-test. The bootstrap test of equality fails to
reject against both alternatives with p-value $0.64\pm0.03$ against
the two-sided and $0.25\pm0.027$ against the one-sided.
In contrast, \citet{wright2013barley} presents
data from \citet{immer1934barley} on barley yields from ten varieties grown at six sites
over two study years.
This is another fully crossed design with larger sample size $10\times 6\times2 = 120$ and different balance.
The estimated variance components are $0.12$ for variety and $1.34$ for site and are both significantly different from zero at the $5\%$ level by an F-test, and the new bootstrap test of equality rejects against both alternatives with p-values of $0.011\pm0.0066$ and $0.005\pm0.0045$ against the two- and one-sided alternatives.

\section{Further simulation results}

\subsection{Nested design, balanced}

\begin{figure}
\centering
\includegraphics[width=6in, height=7in]{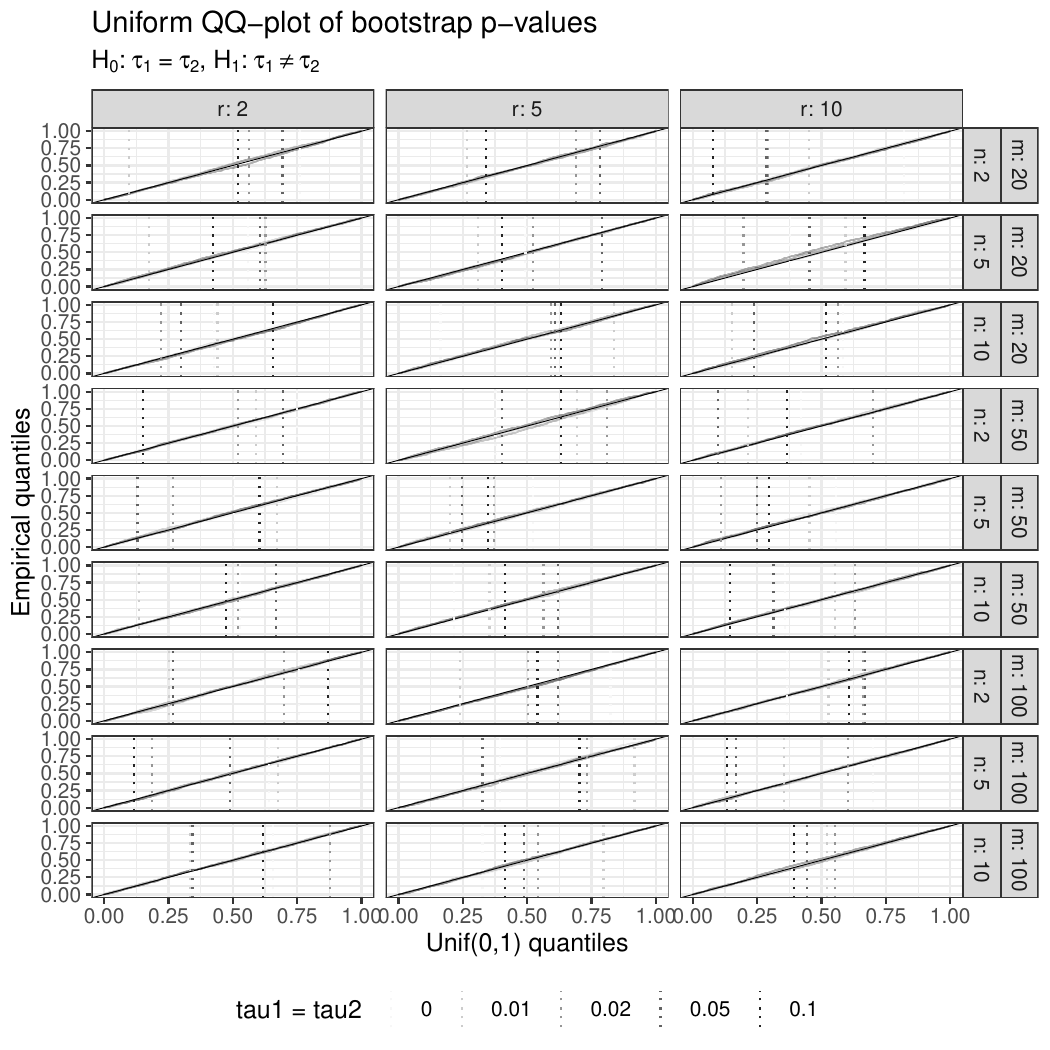}
\caption{Empirical p-values for testing $H_0: \tau_1=\tau_2$ against the two-sided alternative $H_1:\tau_1\neq\tau_2$ in $1000$ simulated datasets from the nested model for each choice of the three sample sizes $m$, $n$, and $r$ and common $\tau$ values in a balanced design. 
The null hypothesis is true
in each simulated dataset and the common $\tau_1=\tau_2$ values
are indicated by separate lines.
Vertical lines indicate the location of maximum dsicrepancy between empirical and theoretical distributions.
}
\label{fig:nestedpvaltwoside}
\end{figure}

\begin{figure}
\centering
\includegraphics[width=6in, height=7in]{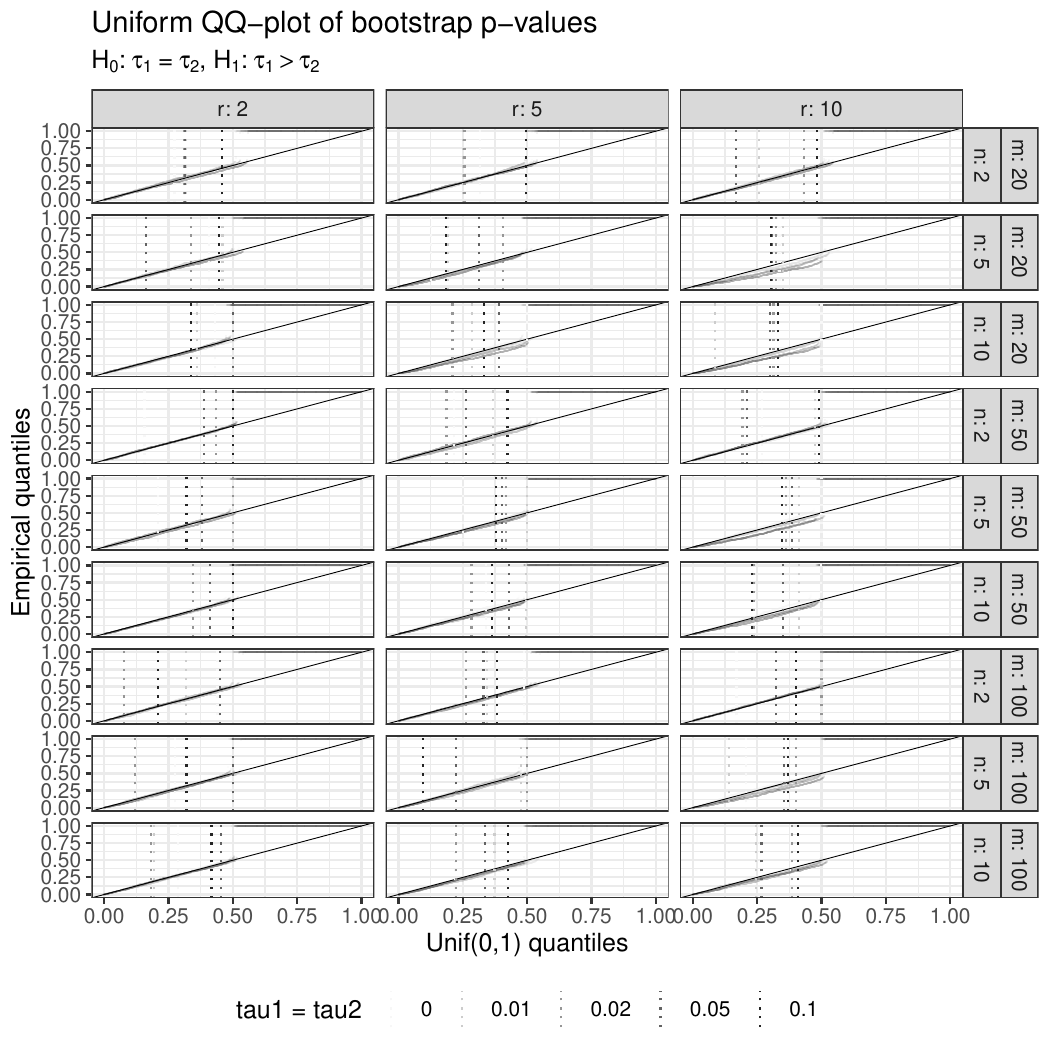}
\caption{Empirical p-values for testing $H_0: \tau_1=\tau_2$ against the one-sided alternative $H_1:\tau_1>\tau_2$ in $1000$ simulated datasets from the nested model for each choice of the three sample sizes $m$, $n$, and $r$ and common $\tau$ values in a balanced design. 
The null hypothesis is true
in each simulated dataset and the common $\tau_1=\tau_2$ values
are indicated by separate lines.
Vertical lines indicate the location of maximum dsicrepancy between empirical and theoretical distributions.
}
\label{fig:nestedpvaloneside}
\end{figure}

We repeat the simulation from Section 6.1 for a nested model, instead using a balanced design consisting of $m$
blocks of $n$ plots each with $r$ replications.
\cref{fig:nestedpvaltwoside,fig:nestedpvaloneside} show QQ-plots of the simulated $p$-values against a $\text{Unif}(0,1)$ distribution.
Each simluation consists of $S=1000$ simulated bootstrapped $p$-values---each itself computed from $B=300$ bootstrap samples---with data 
drawn from a nested model with one combination of $m$, $n$, $r$,
and $\tau_1 = \tau_2$.
The null hypothesis $H_0:\tau_1 = \tau_2$ is true in all cases.
The two-sided $p$-values are expected to fall along the diagonal line, and the one-sided $p$-values are expected to fall
along the diagonal line for $0\leq p\leq0.5$ and then equal $1$.

\begin{figure}
\centering
\includegraphics[width=6in, height=7in]{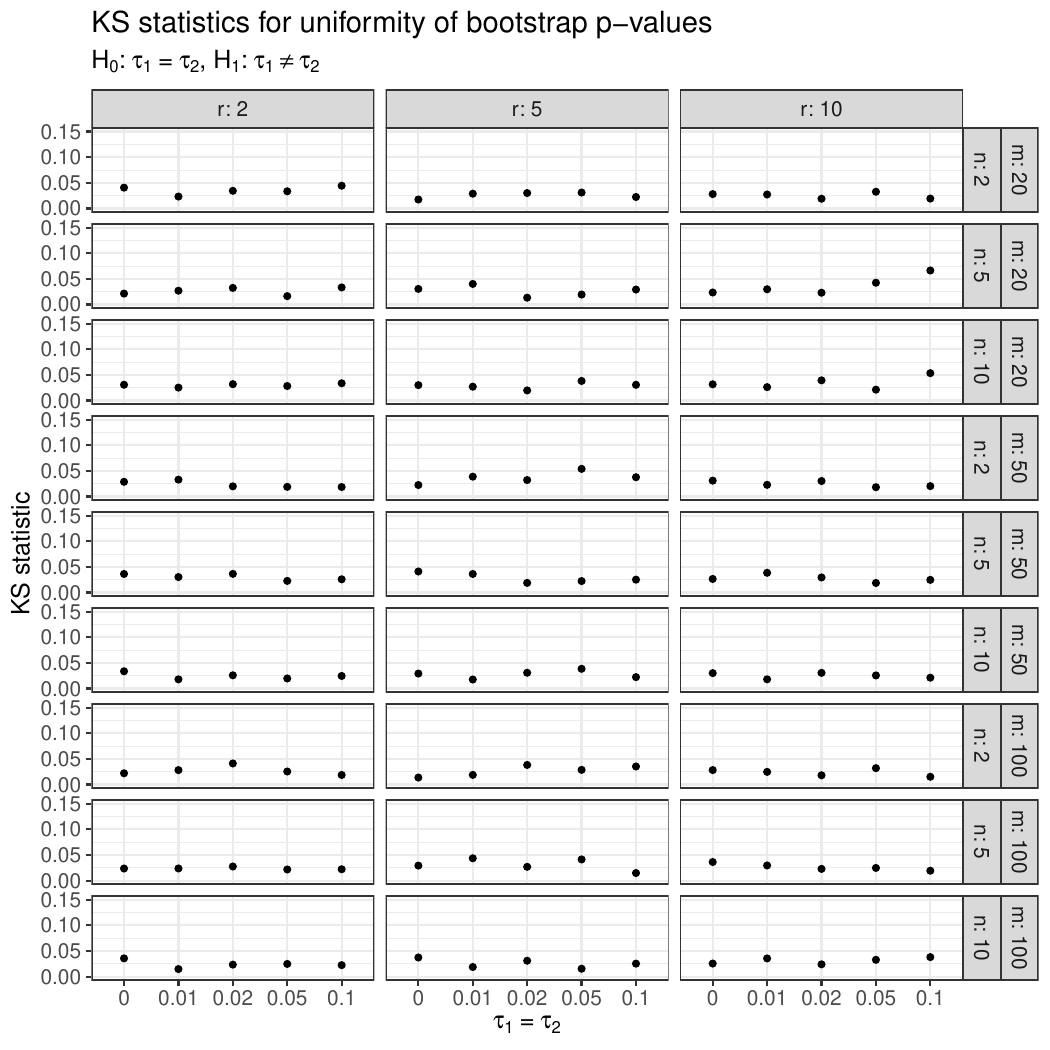}
\caption{KS test statistics for uniformity of the empirical p-values for testing $H_0: \tau_1=\tau_2$ against the two-sided alternative $H_1:\tau_1\neq\tau_2$ in $1000$ simulated datasets from the nested model for each choice of the three sample sizes $m$, $n$, and $r$ and common $\tau$ values in a balanced design. 
The null hypothesis is true in each simulated dataset.
}
\label{fig:nestedkstwoside}
\end{figure}

The majority of $p$-value quantiles are overlapping and all simulations are showing tests that do not appear to deviate
systematically from the uniform null distribution, indicating no empirical evidence of violations of the assumed null
distribution of the test.
\cref{fig:nestedkstwoside} shows the KS statistics, defined as the maximum absolute difference between theoretical and empirical cumulative distribution functions, computed on a fine grid. 
The observed KS statistics are all quite small and show no clear patterns with $m$, $n$, $r$, or $\varcomp$.

\begin{figure}
\centering
\includegraphics[width=6in, height=7in]{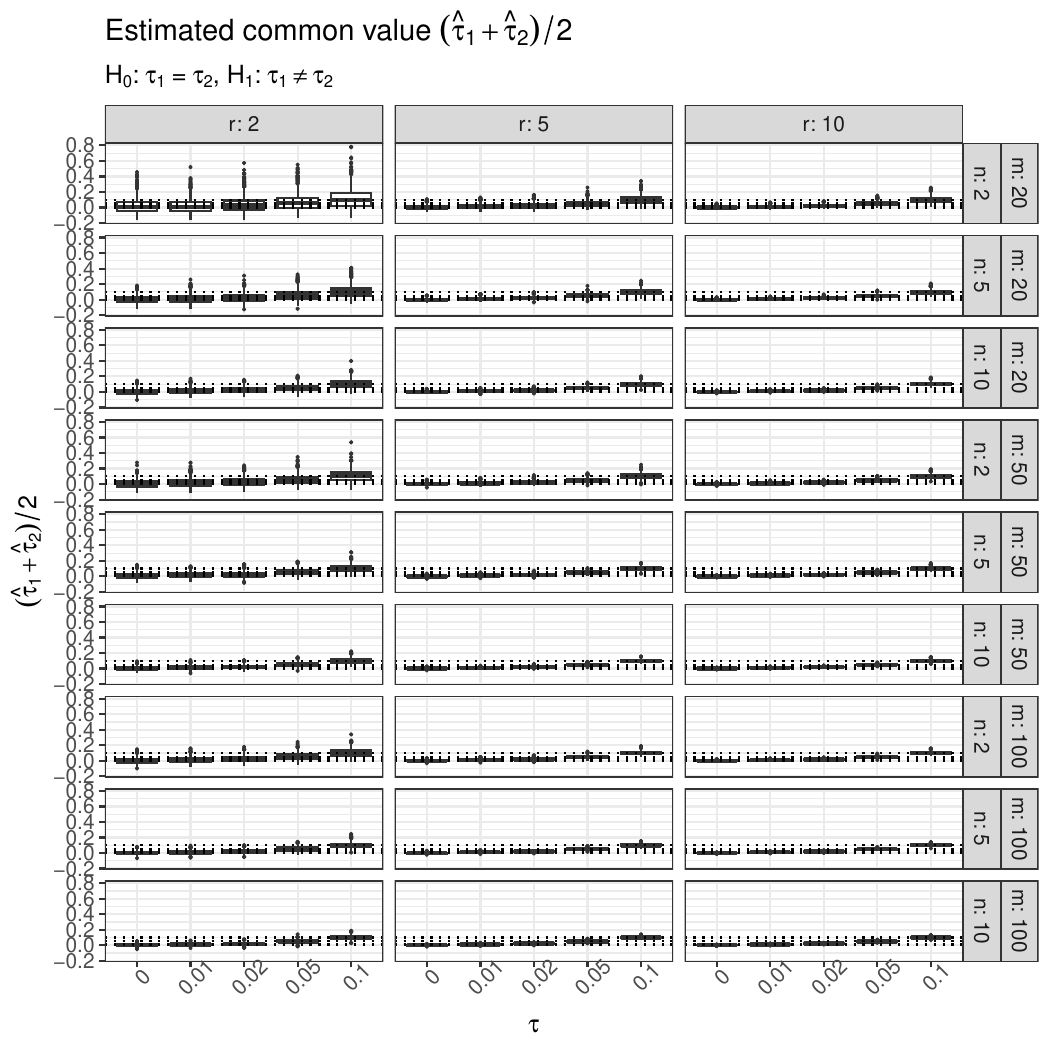}
\caption{Estimated common value $\mb{Q}_2\Tr\widehat{\varcomp}=(1/2)(\widehat{\tau}_1 + \widehat{\tau}_2)$ in $1000$ simulated datasets from the nested model for each choice of the three sample sizes $m$, $n$, and $r$ and common $\tau$ values in a balanced design. 
The null hypothesis of $\tau_1=\tau_2$ is true in each simulated dataset.
Larger, more balanced data tend to result in less uncertainty.
Horizontal lines show the true value of $\tau_1=\tau_2$
for each simulation.
}
\label{fig:nestedbalancedtau}
\end{figure}

\begin{figure}
\centering
\includegraphics[width=6in, height=7in]{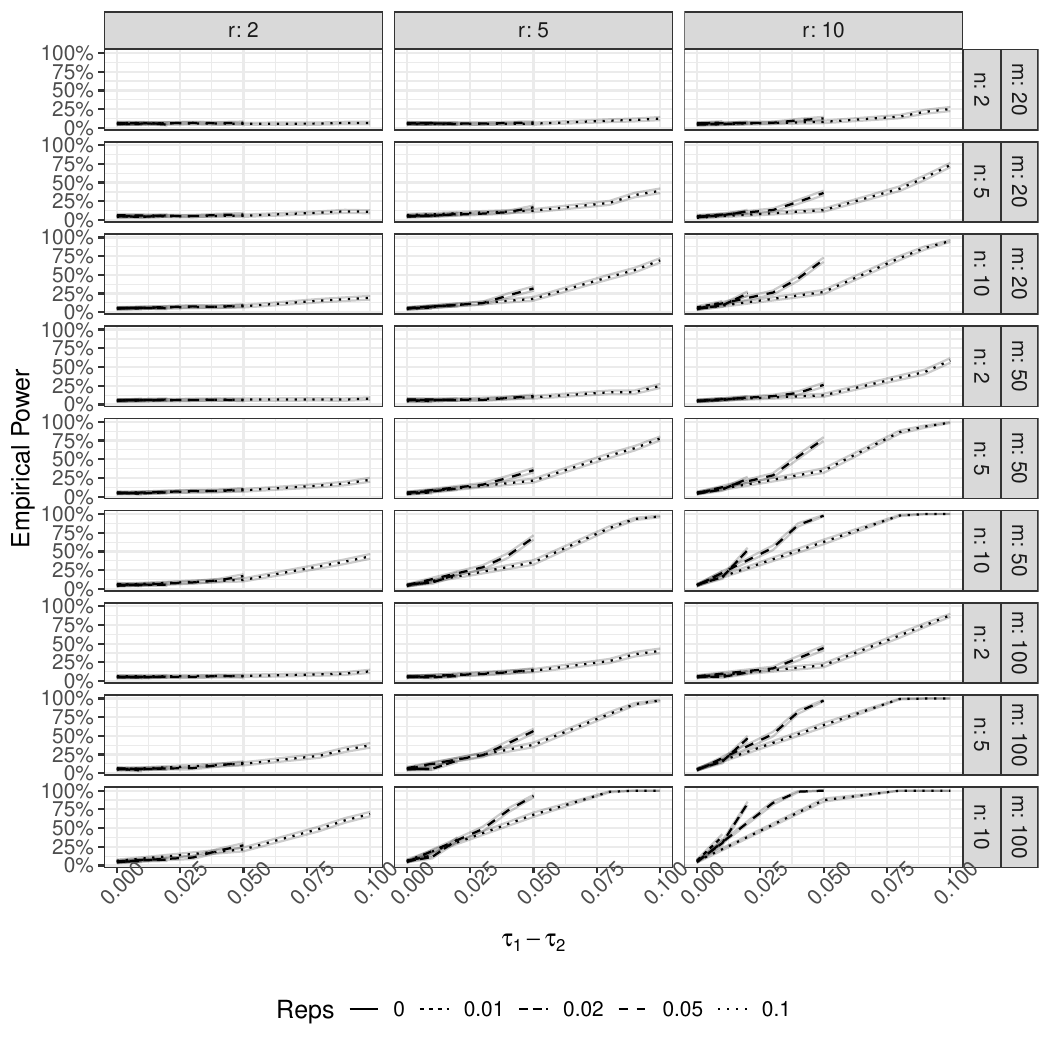}
\caption{Empirical power (with Monte Carlo standard error bands) for testing $H_0: \tau_1=\tau_2$ against the one-sided alternative $H_1:\tau_1>\tau_2$ in $1000$ simulated datasets from the nested model for each choice of the $m$, $n$, and $r$ and $\tau$ in a balanced design. 
Power increases as the sample size increases through any of the 
number of blocks, number of plots, or number of replicates.
Power also increases as the effect size increases.
Each line goes only as far as its $\tau_1$ value.
}
\label{fig:nestedpoweroneside}
\end{figure}

\cref{fig:nestedpoweroneside,fig:nestedpowertwoside} show the empirical power against the one-sided alternative that $\tau_1 > \tau_2$
and the two-sided alternative that $\tau_1=\tau_2$.
Each line corresponds to one fixed value of $\tau_1$ and the $x$-axis corresponds to the difference $\tau_1 - \tau_2$
or $\lvert\tau_1 - \tau_2\rvert$
for each simulated dataset. 
Power increases when any of the samples sizes or the effect size is increased.
The test appears to have very little power with the smallest data sizes of $m=20$ groups with $n=2$ observations per group
and $r=2$ replications per observation, but this increases in a smooth, monotone manner towards having high power for the largest sizes and
effect sizes tested. 

\begin{figure}
\centering
\includegraphics[width=6in, height=7in]{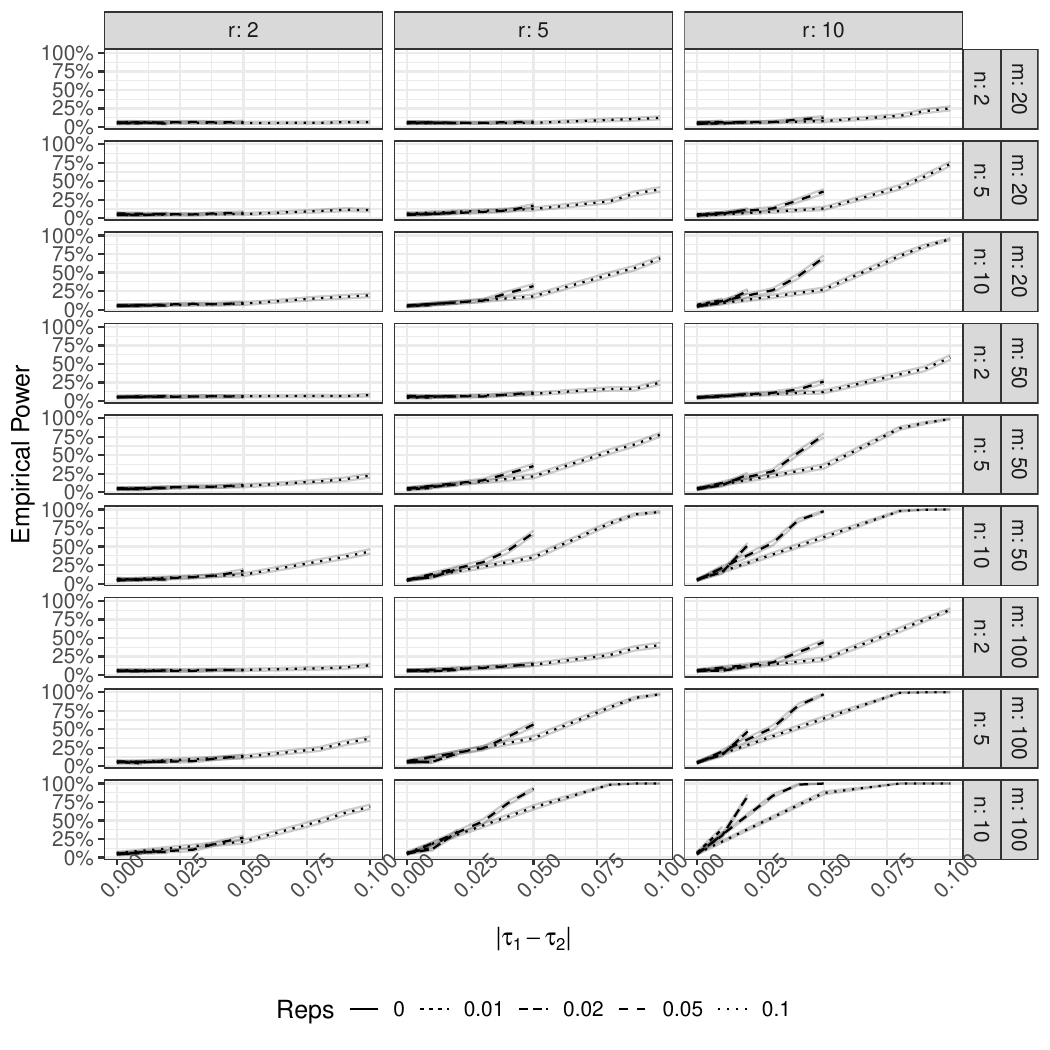}
\caption{Empirical power (with Monte Carlo standard error bands) for testing $H_0: \tau_1=\tau_2$ against the two-sided alternative $H_1:\tau_1\neq\tau_2$ in $1000$ simulated datasets from the nested model for each choice of the $m$, $n$, and $r$ and $\tau$ in a balanced design. 
Power increases as the sample size increases through any of the 
number of blocks, number of plots, or number of replicates.
Power also increases as the effect size increases.
Each line goes only as far as its $\tau_1$ value.
}
\label{fig:nestedpowertwoside}
\end{figure}

\subsection{Nested design, unbalanced}

Simulations from a nested model with unbalanced sample sizes were reported in Section 6.1.
This section contains further details.
As reported in the main text, each simulation consists of $S=1000$ simulated bootstrapped $p$-values---each itself computed from $B=300$ bootstrap samples---with data 
drawn from a nested model with one combination of $m$, $n$, $r$,
and $\tau_1 = \tau_2$.
To achieve imbalance, $m$ groups of size $n_i$ were sampled
where $n_i\sim\text{Unif}(2, 2n-2)$ independently, so that
$\text{min}(n_i) = 2$ and $\mathbb{E}(n_i) = n$. 
Each subgroup consisted of $r_i$ replications
with $r_i\sim\text{Unif}(2, 2r-2)$, again so that
$\text{min}(r_i) = 2$ and $\mathbb{E}(r_i) = r$.
The null hypothesis $H_0:\tau_1 = \tau_2$ is true in all cases.
The two-sided $p$-values are expected to fall along the diagonal line, and the one-sided $p$-values are expected to fall
along the diagonal line for $0\leq p\leq0.5$ and then equal $1$.

\begin{figure}
\centering
\includegraphics[width=6in, height=7in]{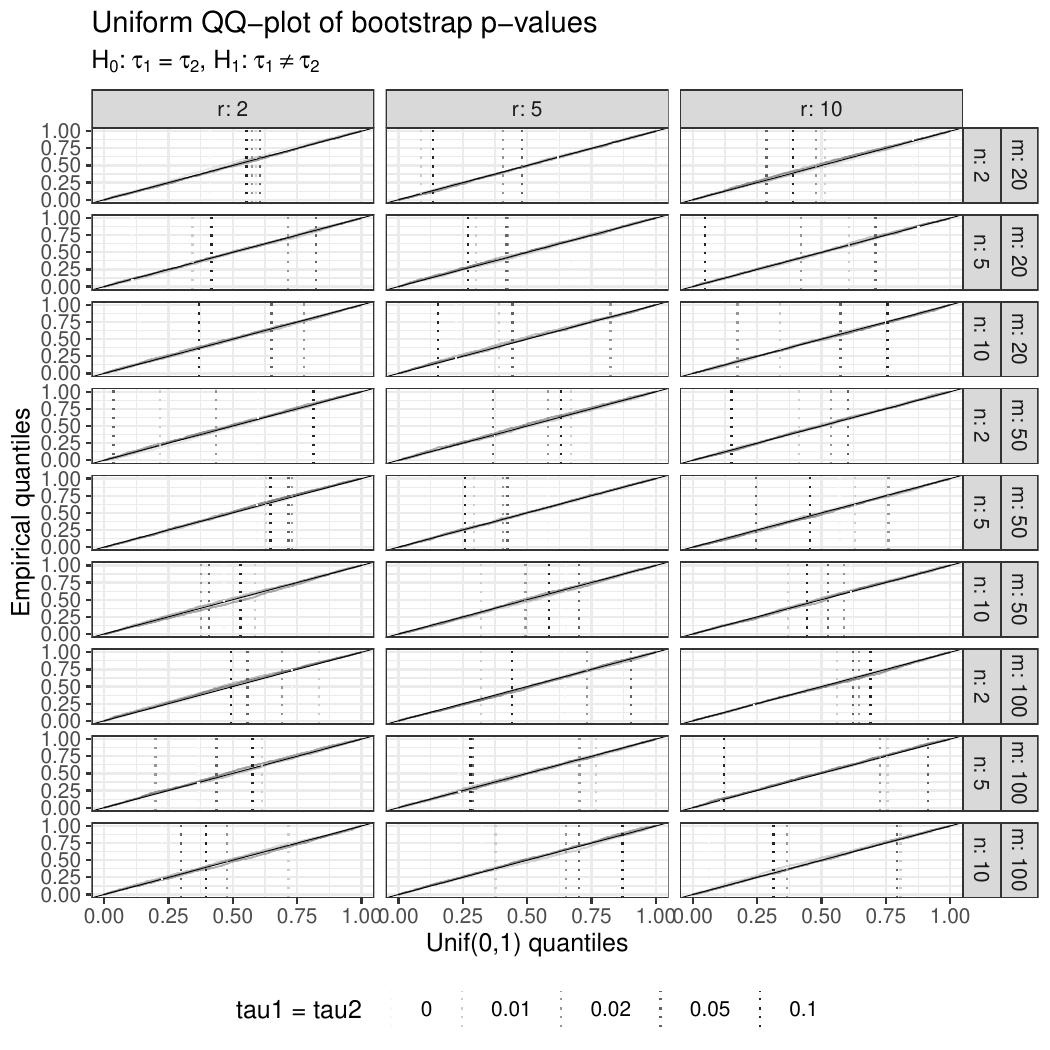}
\caption{Empirical p-values for testing $H_0: \tau_1=\tau_2$ against the two-sided alternative $H_1:\tau_1\neq\tau_2$ in $1000$ simulated datasets from the nested model with an unbalanced design having $m$ groups of average size $n$ and
average number of replications $r$, and common $\tau$ values.
The null hypothesis is true
in each simulated dataset and the common $\tau_1=\tau_2$ values
are indicated by separate lines.
Vertical lines indicate the location of maximum dsicrepancy between empirical and theoretical distributions.
}
\label{fig:nestedpvaltwosideunbalanced}
\end{figure}

\begin{figure}
\centering
\includegraphics[width=6in, height=7in]{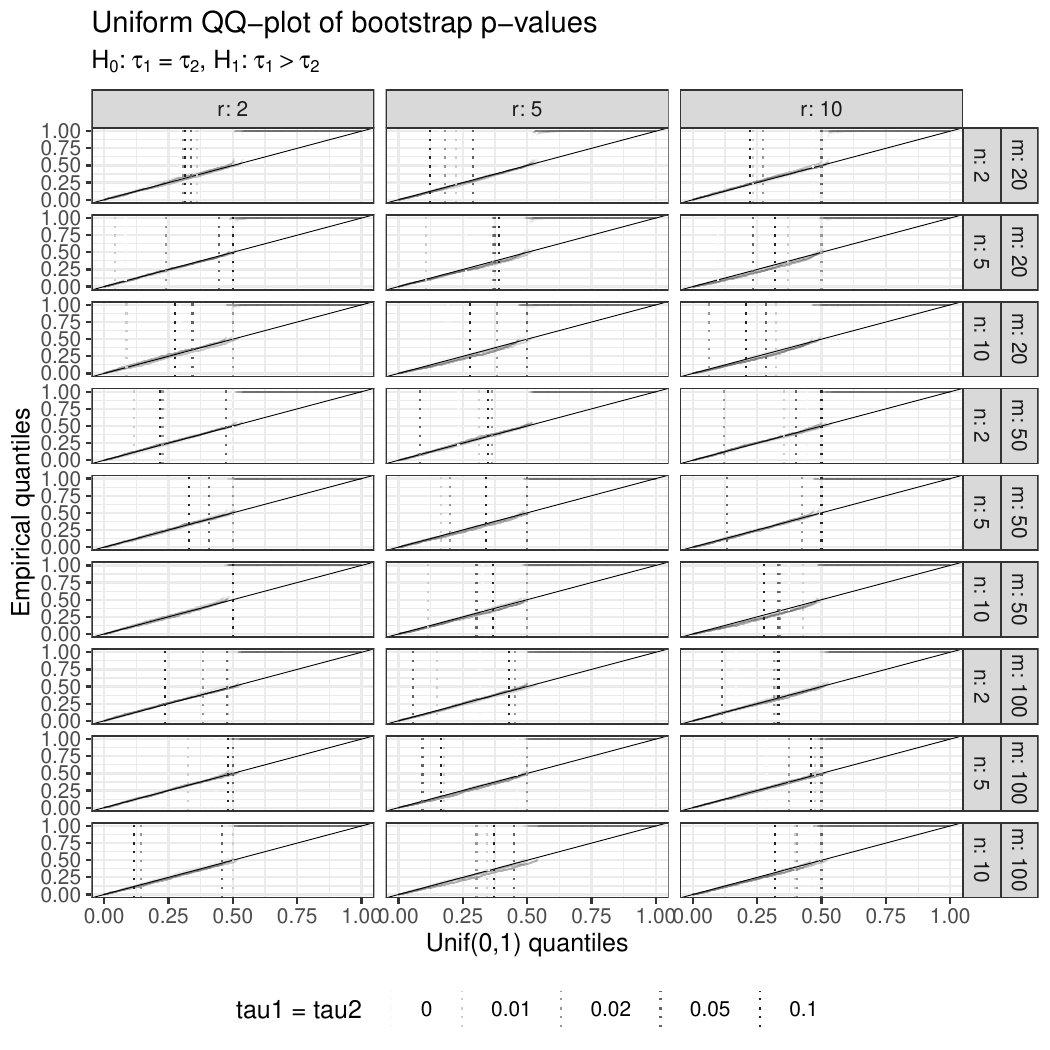}
\caption{Empirical p-values for testing $H_0: \tau_1=\tau_2$ against the one-sided alternative $H_1:\tau_1>\tau_2$ in $1000$ simulated datasets from the nested model with an unbalanced design havin $m$ groups of average size $n$ and
average number of replications $r$, and common $\tau$ values.
The null hypothesis is true
in each simulated dataset and the common $\tau_1=\tau_2$ values
are indicated by separate lines.
Vertical lines indicate the location of maximum dsicrepancy between empirical and theoretical distributions.
}
\label{fig:nestedpvalonesideunbalanced}
\end{figure}

\begin{figure}
\centering
\includegraphics[width=6in, height=7in]{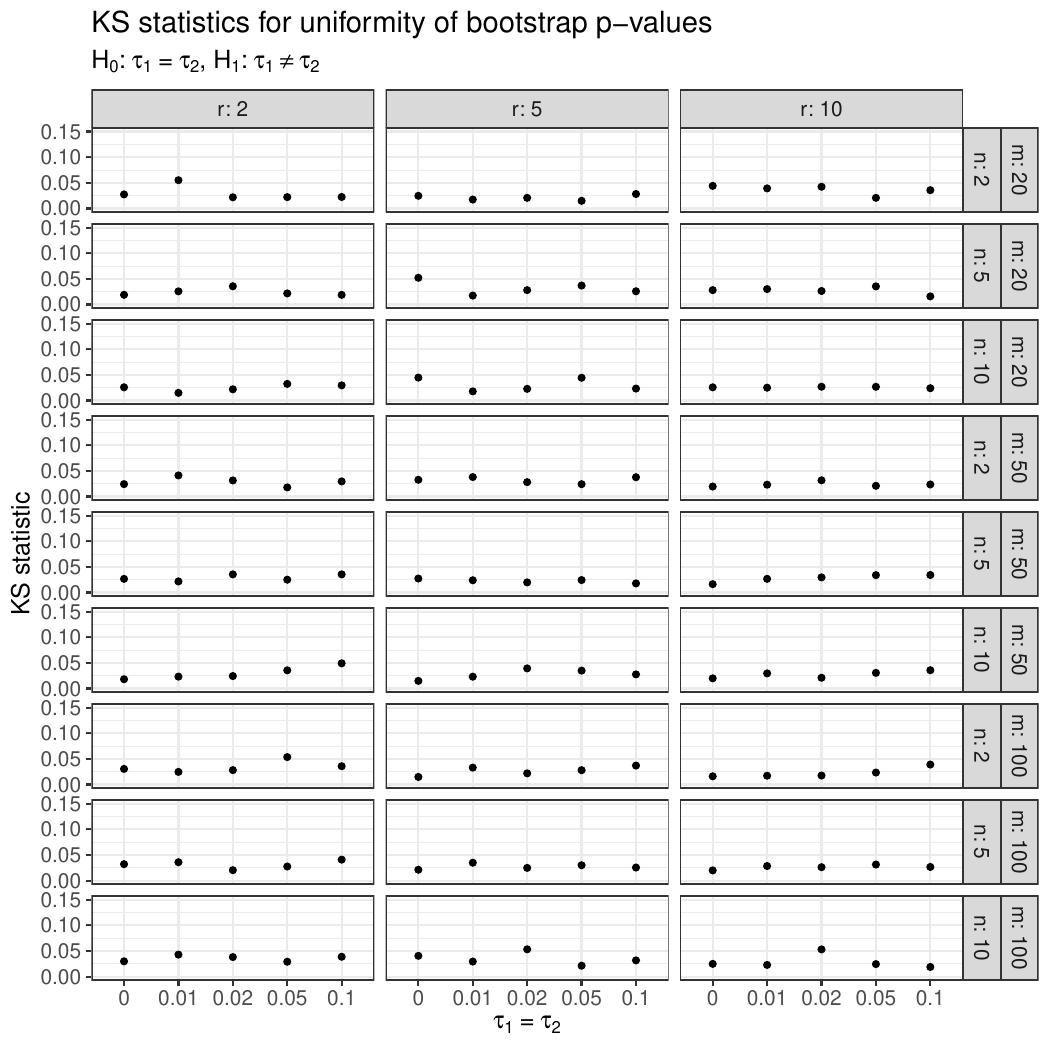}
\caption{KS test statistics for uniformity of the empirical p-values for testing $H_0: \tau_1=\tau_2$ against the two-sided alternative $H_1:\tau_1\neq\tau_2$ in $1000$ simulated datasets from the nested model with an unbalanced design having $m$ groups of average size $n$ and
average number of replications $r$, and common $\tau$ values.
The null hypothesis is true in each simulated dataset.
}
\label{fig:nestedkstwosideunbalanced}
\end{figure}

\cref{fig:nestedpvaltwosideunbalanced,fig:nestedpvalonesideunbalanced} show QQ-plots of the simulated $p$-values against a $\text{Unif}(0,1)$ distribution.
The majority of $p$-value quantiles are overlapping and all simulations are showing tests that do not appear to deviate
systematically from the uniform null distribution, indicating no evidence of empirical violations of the assumed null
distribution of the test.
\cref{fig:nestedkstwosideunbalanced} shows the KS statistics, which are all quite small and with no clear patterns
with $m$, $n$, $r$, or $\varcomp$.

\begin{figure}
\centering
\includegraphics[width=6in, height=7in]{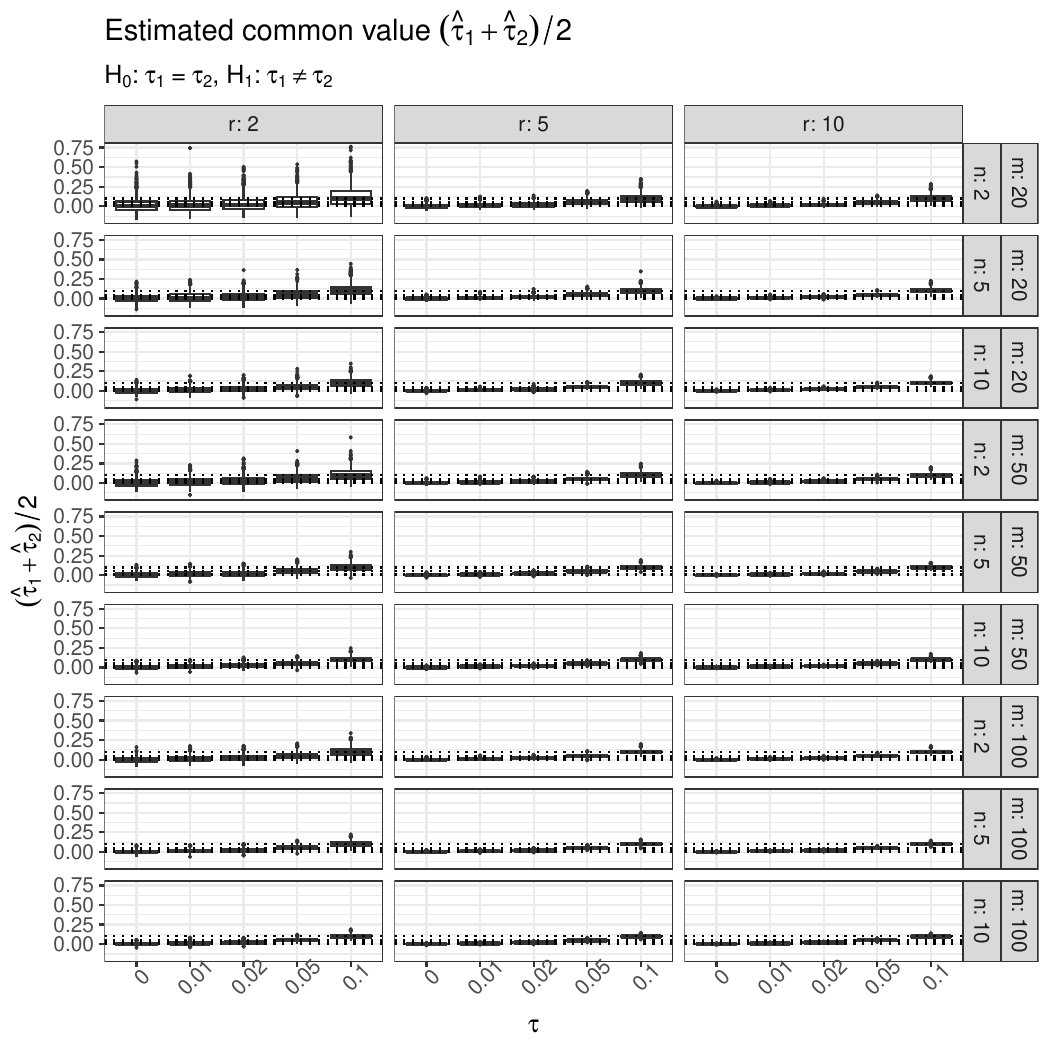}
\caption{Estimated common value $\mb{Q}_2\Tr\widehat{\varcomp}=(1/2)(\widehat{\tau}_1 + \widehat{\tau}_2)$ in $1000$ simulated datasets from the nested model for each choice of the three sample sizes $m$, $n$, and $r$ and common $\tau$ values in an unbalanced design. 
The null hypothesis of $\tau_1=\tau_2$ is true in each simulated dataset.
Larger, more balanced data tend to result in less uncertainty.
Horizontal lines show the true value of $\tau_1=\tau_2$
for each simulation.
}
\label{fig:nestedbalancedtau}
\end{figure}

\begin{figure}
\centering
\includegraphics[width=6in, height=7in]{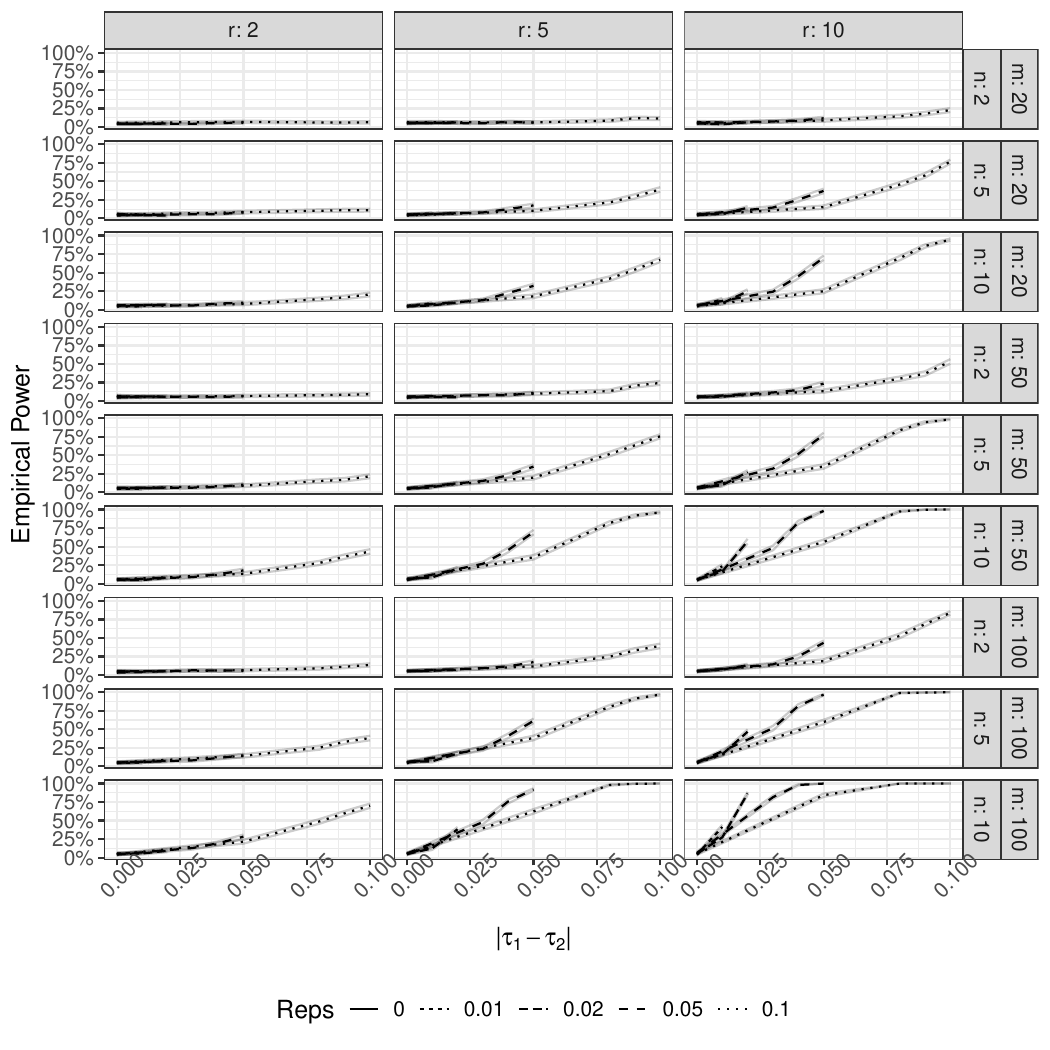}
\caption{Empirical power (with Monte Carlo standard error bands) for testing $H_0: \tau_1=\tau_2$ against the two-sided alternative $H_1:\tau_1\neq\tau_2$ in $1000$ simulated datasets from the nested model with an unbalanced design having $m$ groups of average size $n$ and
average number of replications $r$, and common $\tau$ values.
Power increases as the sample size increases through any of the 
number of blocks, number of plots, or number of replicates.
Power also increases as the effect size increases.
Each line goes only as far as its $\tau_1$ value.
}
\label{fig:nestedpowertwosideunbalanced}
\end{figure}

\begin{figure}
\centering
\includegraphics[width=6in, height=7in]{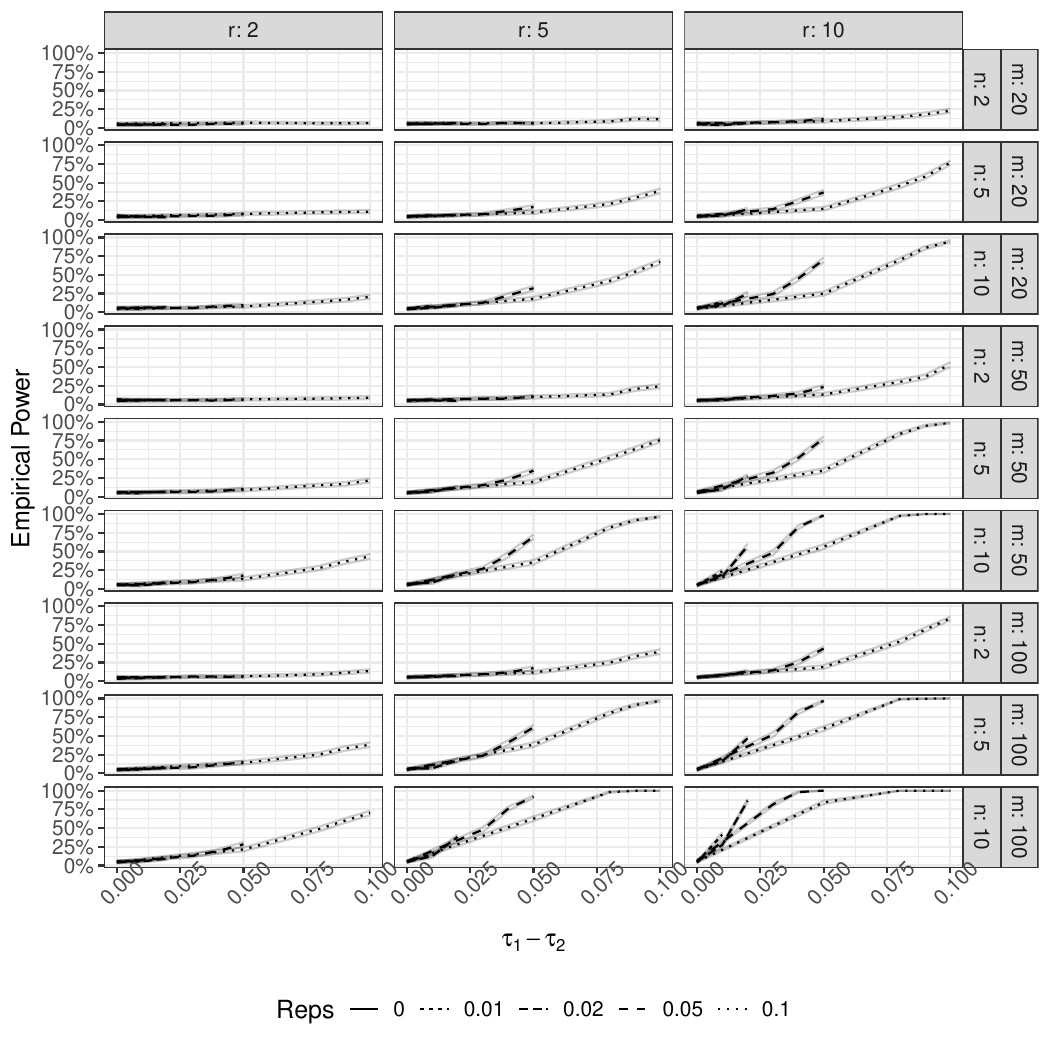}
\caption{Empirical power (with Monte Carlo standard error bands) for testing $H_0: \tau_1=\tau_2$ against the one-sided alternative $H_1:\tau_1>\tau_2$ in $1000$ simulated datasets from the nested model with an unbalanced design having $m$ groups of average size $n$ and
average number of replications $r$, and common $\tau$ values.
Power increases as the sample size increases through any of the 
number of blocks, number of plots, or number of replicates.
Power also increases as the effect size increases.
Each line goes only as far as its $\tau_1$ value.
}
\label{fig:nestedpoweronesideunbalanced}
\end{figure}

\cref{fig:nestedpoweronesideunbalanced,fig:nestedpowertwosideunbalanced} show the empirical power against the one-sided alternative 
that $\tau_1 > \tau_2$ and the two-sided alternative that $\tau_1\neq\tau_2$.
Each line corresponds to one fixed value of $\tau_1$ and the $x$-axis corresponds to the difference $\tau_1 - \tau_2$
for each simulated dataset. 
Power increases when any of the samples sizes or the effect size is increased.
The test appears to have very little power with the smallest data sizes of $m=20$ groups with $n=2$ observations per group
and $r=2$ replications per observation, but this increases in a smooth, monotone manner towards having high power for the largest sizes and
effect sizes tested. 

\subsection{Crossed design}

\begin{figure}
\centering
\includegraphics[width=6in, height=7in]{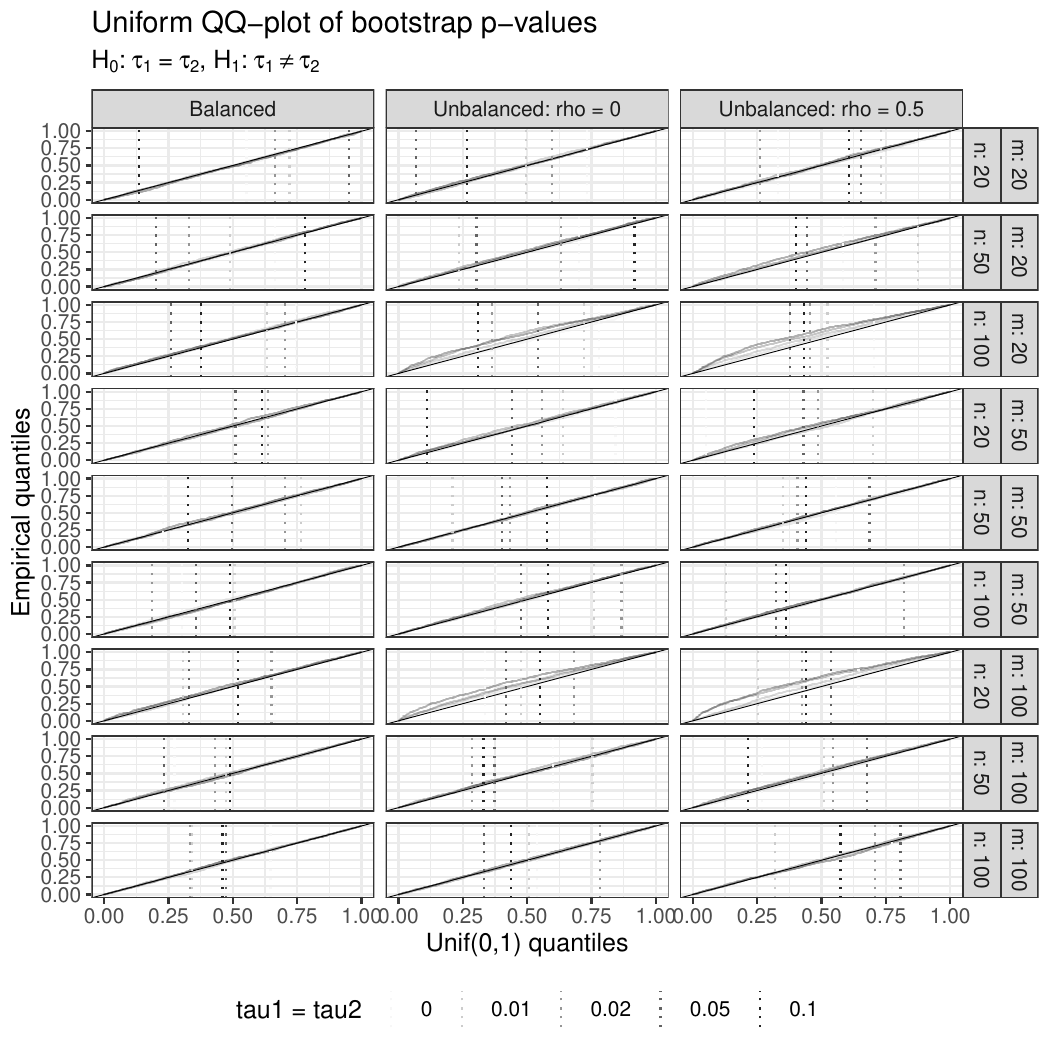}
\caption{Empirical p-values for testing $H_0: \tau_1=\tau_2$ against the two-sided alternative $H_1:\tau_1\neq\tau_2$ in $1000$ simulated datasets from the crossed model for each choice of the two sample sizes $m$ and $n$, common $\tau$ values, and balance parameter $r$.
The null hypothesis is true
in each simulated dataset and the common $\tau_1=\tau_2$ values
are indicated by separate lines.
Vertical lines indicate the location of maximum dsicrepancy between empirical and theoretical distributions.
}
\label{fig:crossedpvaltwoside}
\end{figure}

\begin{figure}
\centering
\includegraphics[width=6in, height=7in]{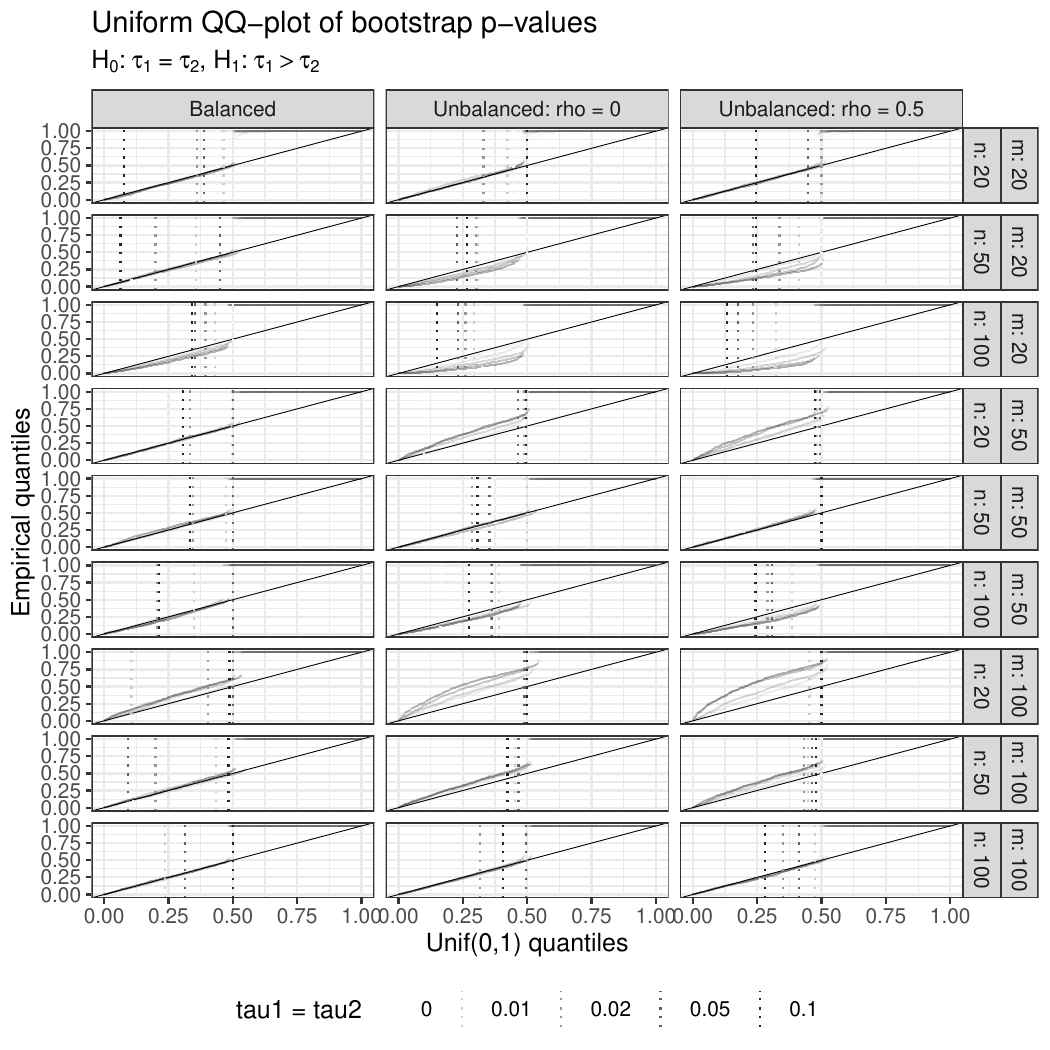}
\caption{Empirical p-values for testing $H_0: \tau_1=\tau_2$ against the one-sided alternative $H_1:\tau_1>\tau_2$ in $1000$ simulated datasets from the crossed model for each choice of the two sample sizes $m$ and $n$, common $\tau$ values, and balance parameter $r$.
The null hypothesis is true
in each simulated dataset and the common $\tau_1=\tau_2$ values
are indicated by separate lines.
Vertical lines indicate the location of maximum dsicrepancy between empirical and theoretical distributions.}
\label{fig:crossedpvaloneside}
\end{figure}

Simulations from a crossed model with balanced and unbalanced sample sizes were reported in Section 6.2.
This section contains further details.
\cref{fig:crossedpvaltwoside,fig:crossedpvaloneside} show QQ-plots of the simulated $p$-values against a $\text{Unif}(0,1)$ distribution.
Like in the nested simulations and as reported in the main text, each simluation consists 
of $S=1000$ simulated bootstrapped $p$-values each computed from $B=300$ bootstrap samples, but with data 
drawn from a crossed model with one combination of $m$, $n$,
and $\tau_1 = \tau_2$.
Here the $r$ values indicate the degree of balanced: $r=-1$ indicates perfect balance, $r=0$ indicates randomly sampled
pairs $(i,j)$ that generate a balanced design on average, and $r>0$ indicates a lack of balance with larger $r$ indicating greater
imbalance.
The null hypothesis $H_0:\tau_1 = \tau_2$ is true in all cases.

\begin{figure}
\centering
\includegraphics[width=6in, height=7in]{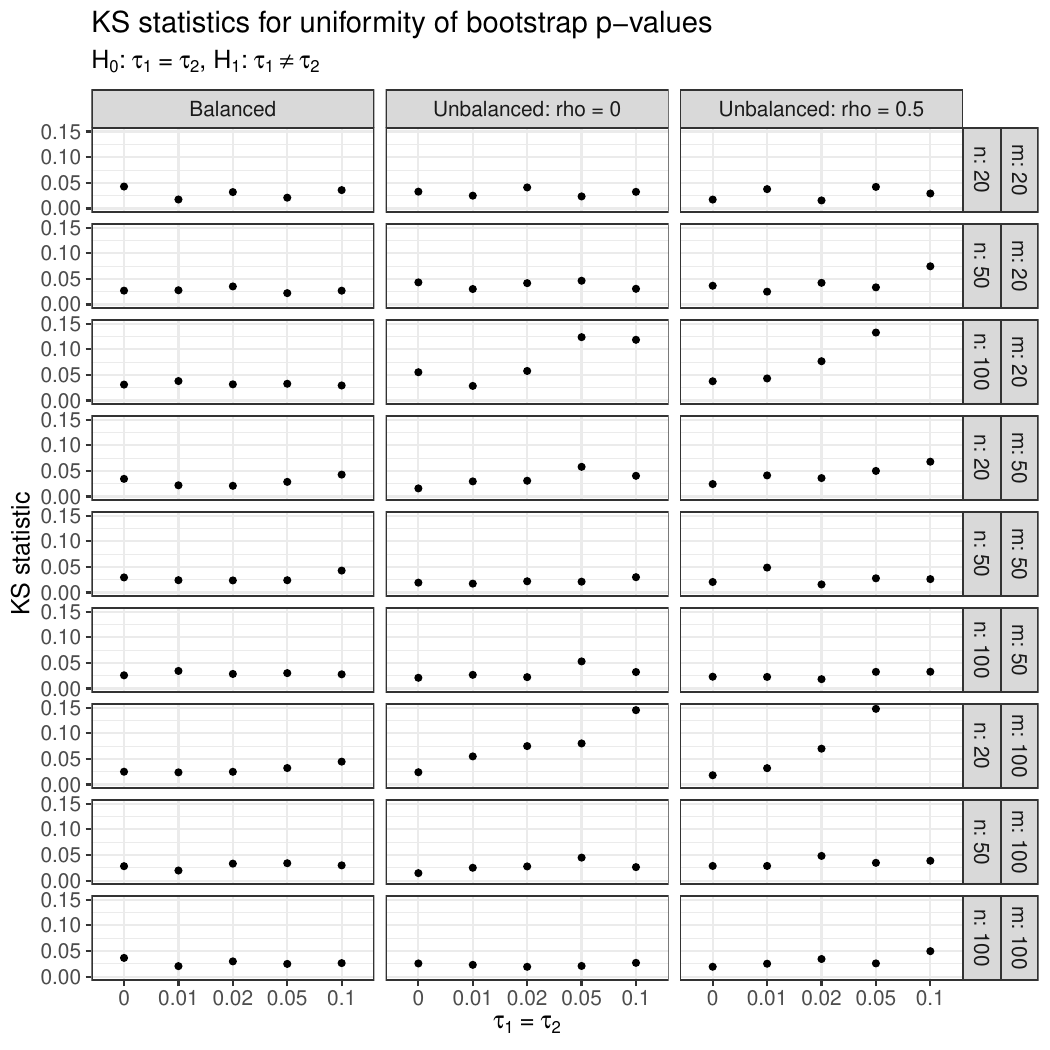}
\caption{KS test statistics for uniformity of the empirical p-values for testing $H_0: \tau_1=\tau_2$ against the two-sided alternative $H_1:\tau_1\neq\tau_2$ in $1000$ simulated datasets from the crossed model for each choice of the two sample sizes $m$ and $n$, common $\tau$ values, and balance parameter $r$.
The null hypothesis is true in each simulated dataset.
}
\label{fig:crossedkstwoside}
\end{figure}

\begin{figure}
\centering
\includegraphics[width=6in, height=7in]{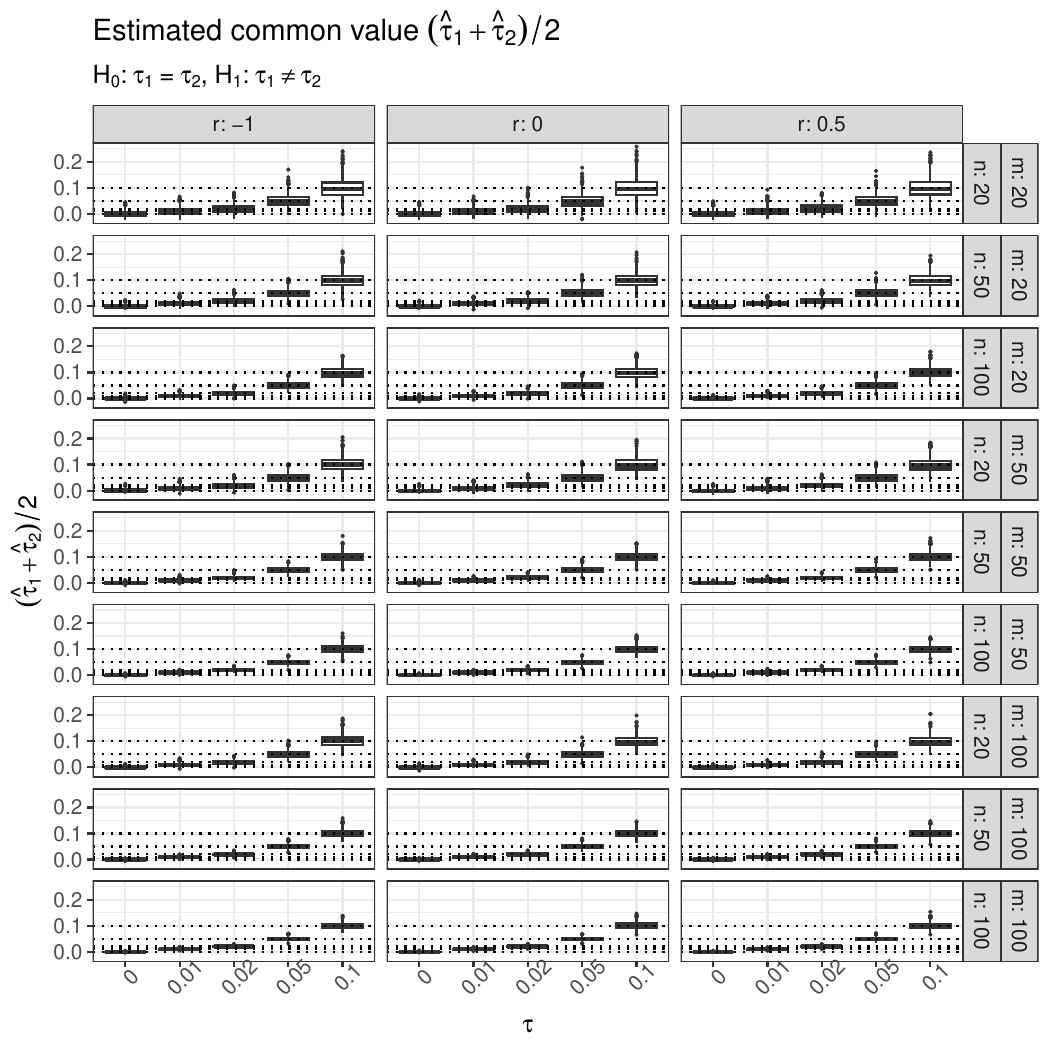}
\caption{Estimated common value $\mb{Q}_2\Tr\widehat{\varcomp}=(1/2)(\widehat{\tau}_1 + \widehat{\tau}_2)$ in $1000$ simulated datasets from the crossed model for each choice of the two sample sizes $m$ and $n$, common $\tau$ values, and balance parameter $r$.
The null hypothesis is true in each simulated dataset.
}
\label{fig:crossedtau}
\end{figure}

\cref{fig:crossedkstwoside} shows most KS statistics being small, but large KS statistics for the most unbalanced designs having
$m=100$ and $n=20$ or vice-versa and being unbalanced with $r=0,0.5$.
Figure \ref{fig:crossedpvaltwoside} indicates that the discrepancies for the two-sided are all in the direction that indicates the test is too conservative,
which is the better of the two possible errors although still concerning.
Figure \ref{fig:crossedpvaloneside} shows examples of the test being conservative and too optimistic.
\cref{fig:crossedtau} shows the variability in estimated
common values $\mb{Q}_2\Tr\widehat{\varcomp}$ across
simulations, with some increased variability for the cases in which the tests have incorrect size.

\begin{figure}
\centering
\includegraphics[width=6in, height=7.7in]{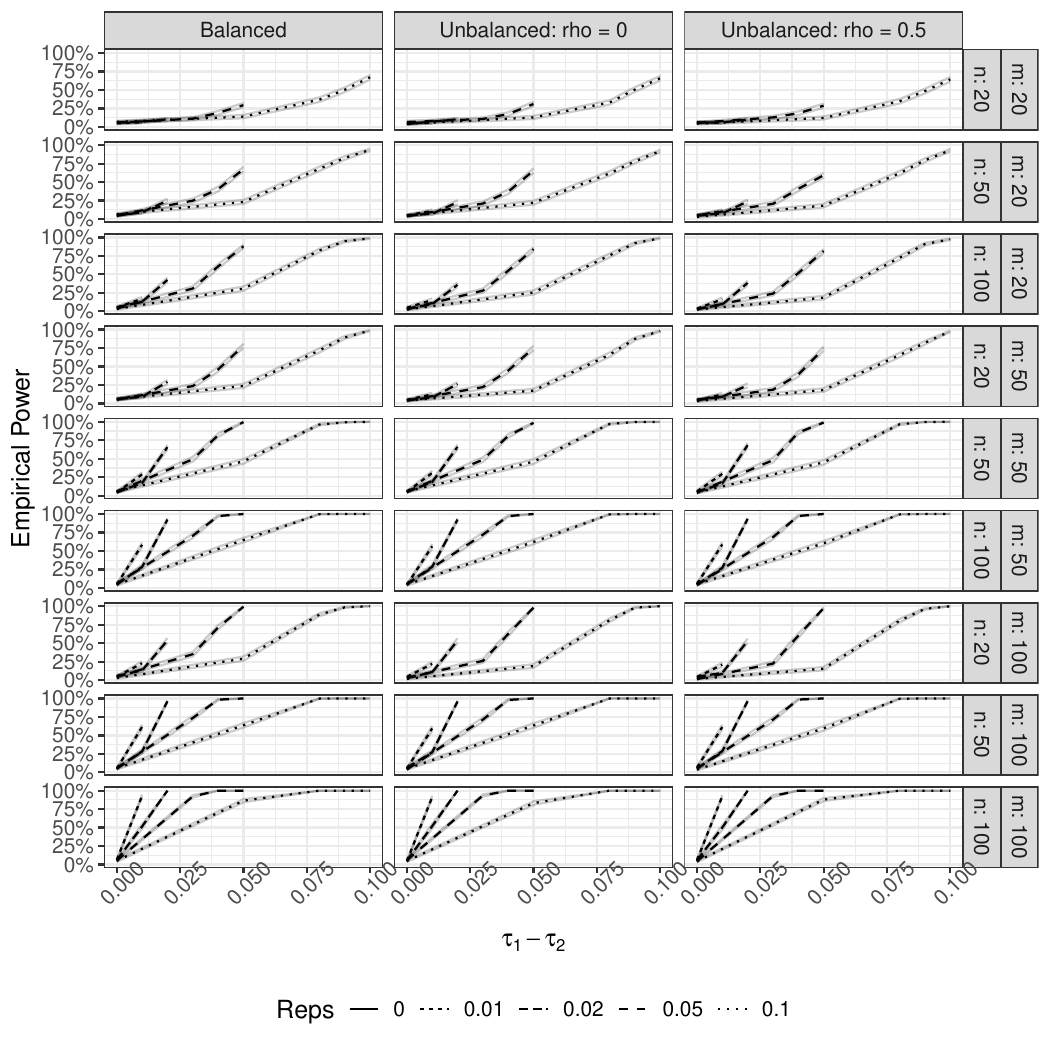}
\caption{Empirical power for testing $H_0: \tau_1=\tau_2$ against the one-sided alternative $H_1:\tau_1>\tau_2$ in $1000$ simulated datasets from the crossed model for each choice of the two sample sizes $m$ and $n$ and common $\tau$ values and for each choice of balance parameter $r=-1, 0, 0.5$. In the designs with more unbalanced sample sizes the power decreases for less balanced effects.
}
\label{fig:crossedpoweroneside}
\end{figure}

\cref{fig:crossedpoweroneside} shows the empirical power against the one-sided alternative that $\tau_1 > \tau_2$.
Each line corresponds to one fixed value of $\tau_1$ and the $x$-axis corresponds to the difference $\tau_1 - \tau_2$
for each simulated dataset.
Power increases when either of the samples sizes or the effect size is increased.
The test appears to have moderate power with the smallest data sizes of $m=n=20$ levels, but this increases in a smooth, 
monotone manner towards having high power for the largest sizes and effect sizes tested. 
The test appears to have good power, however in the
very unbalanced cases this comes at the expense of 
correct size.

\bibliographystyle{apalike}

\bibliography{references}

@article{hui2019testing,
  title={Testing random effects in linear mixed models: another look at the F-test (with discussion)},
  author={Hui, FKC and M{\"u}ller, Samuel and Welsh, AH},
  journal={Australian \& New Zealand Journal of Statistics},
  volume={61},
  number={1},
  pages={61--84},
  year={2019},
  publisher={Wiley Online Library}
}

@article{crainiceanu2004likelihood,
  title={Likelihood ratio tests in linear mixed models with one variance component},
  author={Crainiceanu, Ciprian M and Ruppert, David},
  journal={Journal of the Royal Statistical Society Series B: Statistical Methodology},
  volume={66},
  number={1},
  pages={165--185},
  year={2004},
  publisher={Oxford University Press}
}

@article{wang2012testing,
  title={On testing an unspecified function through a linear mixed effects model with multiple variance components},
  author={Wang, Yuanjia and Chen, Huaihou},
  journal={Biometrics},
  volume={68},
  number={4},
  pages={1113--1125},
  year={2012},
  publisher={Oxford University Press}
}

@article{claeskens2004restricted,
  title={Restricted likelihood ratio lack-of-fit tests using mixed spline models},
  author={Claeskens, Gerda},
  journal={Journal of the Royal Statistical Society Series B: Statistical Methodology},
  volume={66},
  number={4},
  pages={909--926},
  year={2004},
  publisher={Oxford University Press}
}

@article{greven2008restricted,
  title={Restricted likelihood ratio testing for zero variance components in linear mixed models},
  author={Greven, Sonja and Crainiceanu, Ciprian M and K{\"u}chenhoff, Helmut and Peters, Annette},
  journal={Journal of Computational and Graphical Statistics},
  volume={17},
  number={4},
  pages={870--891},
  year={2008},
  publisher={Taylor \& Francis}
}

@article{wood2013simple,
  title={A simple test for random effects in regression models},
  author={Wood, Simon N},
  journal={Biometrika},
  volume={100},
  number={4},
  pages={1005--1010},
  year={2013},
  publisher={Oxford University Press}
}

@article{stram1994variance,
  title={Variance components testing in the longitudinal mixed effects model},
  author={Stram, Daniel O and Lee, Jae Won},
  journal={Biometrics},
  pages={1171--1177},
  year={1994},
  publisher={JSTOR}
}

@article{giampaoli2009likelihood,
  title={Likelihood ratio tests for variance components in linear mixed models},
  author={Giampaoli, Viviana and Singer, Julio M},
  journal={Journal of Statistical Planning and Inference},
  volume={139},
  number={4},
  pages={1435--1448},
  year={2009},
  publisher={Elsevier}
}

@article{crainiceanu2005exact,
  title={Exact likelihood ratio tests for penalised splines},
  author={Crainiceanu, Ciprian and Ruppert, David and Claeskens, Gerda and Wand, Matthew P},
  journal={Biometrika},
  volume={92},
  number={1},
  pages={91--103},
  year={2005},
  publisher={Oxford University Press}
}

@article{verbeke2003use,
  title={The use of score tests for inference on variance components},
  author={Verbeke, Geert and Molenberghs, Geert},
  journal={Biometrics},
  volume={59},
  number={2},
  pages={254--262},
  year={2003},
  publisher={Oxford University Press}
}

@article{qu2013linear,
  title={Linear score tests for variance components in linear mixed models and applications to genetic association studies},
  author={Qu, Long and Guennel, Tobias and Marshall, Scott L},
  journal={Biometrics},
  volume={69},
  number={4},
  pages={883--892},
  year={2013},
  publisher={Oxford University Press}
}

@article{battey2024anomaly,
  title={An anomaly arising in the analysis of processes with more than one source of variability},
  author={Battey, HS and McCullagh, Peter},
  journal={Biometrika},
  volume={111},
  number={2},
  pages={677--689},
  year={2024},
  publisher={Oxford University Press}
}

@article{zhang2024fast,
  title={Fast and reliable confidence intervals for a variance component},
  author={Zhang, Yiqiao and Ekvall, Karl Oskar and Molstad, Aaron J},
  journal={Biometrika},
  volume={112},
  number={2},
  pages={asaf010},
  year={2025},
  publisher={Oxford University Press}
}

@article{bates2015fitting,
  title={Fitting linear mixed-effects models using lme4},
  author={Bates, Douglas and Maechler, Martin and Bolker, Ben and Walker, Steve},
  journal={Journal of Statistical Software},
  volume={67},
  number={1},
  pages={1--48},
  year={2015}
}

@article{patterson1971recovery,
  title={Recovery of inter-block information when block sizes are unequal},
  author={Patterson, H Desmond and Thompson, Robin},
  journal={Biometrika},
  volume={58},
  number={3},
  pages={545--554},
  year={1971},
  publisher={Oxford University Press}
}

@article{laird1982random,
  title={Random-effects models for longitudinal data},
  author={Laird, Nan M and Ware, James H},
  journal={Biometrics},
  pages={963--974},
  year={1982},
  publisher={JSTOR}
}

@incollection{venables2002random,
  title={Random and mixed effects},
  author={Venables, William N and Ripley, Brian D},
  booktitle={Modern applied statistics with S-PLUS},
  pages={297--322},
  year={2002},
  publisher={Springer}
}

@article{yates1935complex,
  title={Complex experiments},
  author={Yates, Frank},
  journal={Supplement to the Journal of the Royal Statistical Society},
  volume={2},
  number={2},
  pages={181--247},
  year={1935},
  publisher={JSTOR}
}

@book{davies1947statistical,
  title={Statistical methods in research and production.},
  author={Davies, Owen L},
  year={1947},
  publisher={Oliver and Boyd}
}

@book{pinheiro2000mixed,
  title={Mixed-effects models in S and S-PLUS},
  author={Pinheiro, Jos{\'e} C and Bates, Douglas M},
  year={2000},
  publisher={Springer}
}

@book{davies1972statistical,
  title={Statistical Methods in Research and Production},
  author={O.L. Davies and P.L. Goldsmith},
  year={1972},
  publisher={Oliver and Boyd}
}

@article{immer1934barley,
  title={Statistical Determination of Barley Varietal Adaptation},
  author={R. F. Immer and H. K. Hayes and LeRoy Powers},
  journal={Journal of the American Society of Agronomy},
  volume={26},
  number={},
  pages={403--419},
  year={1934},
  publisher={}
}

@article{wright2013barley,
  title={Revisiting Immer's Barley Data},
  author={Kevin Wright},
  journal={The American Statistician},
  volume={67},
  number={3},
  pages={129--133},
  year={2013},
  publisher={}
}

@article{chen2008algorithm,
  title={Algorithm 887: CHOLMOD, supernodal sparse Cholesky factorization and update/downdate},
  author={Chen, Yanqing and Davis, Timothy A and Hager, William W and Rajamanickam, Sivasankaran},
  journal={ACM Transactions on Mathematical Software (TOMS)},
  volume={35},
  number={3},
  pages={1--14},
  year={2008},
  publisher={ACM New York, NY, USA}
}

@article{bezanson2017julia,
  title={Julia: A fresh approach to numerical computing},
  author={Bezanson, Jeff and Edelman, Alan and Karpinski, Stefan and Shah, Viral B},
  journal={SIAM review},
  volume={59},
  number={1},
  pages={65--98},
  year={2017},
  publisher={SIAM}
}

@article{hoffman2016variancepartition,
  title={variancePartition: interpreting drivers of variation in complex gene expression studies},
  author={Hoffman, Gabriel E and Schadt, Eric E},
  journal={BMC bioinformatics},
  volume={17},
  number={1},
  pages={483},
  year={2016},
  publisher={Springer}
}

@article{nye2004large,
  title={How large are teacher effects?},
  author={Nye, Barbara and Konstantopoulos, Spyros and Hedges, Larry V},
  journal={Educational evaluation and policy analysis},
  volume={26},
  number={3},
  pages={237--257},
  year={2004},
  publisher={Sage Publications Sage CA: Los Angeles, CA}
}

@article{messier2010traits,
  title={How do traits vary across ecological scales? A case for trait-based ecology},
  author={Messier, Julie and McGill, Brian J and Lechowicz, Martin J},
  journal={Ecology letters},
  volume={13},
  number={7},
  pages={838--848},
  year={2010},
  publisher={Wiley Online Library}
}

@article{hill2008data,
  title={Data and theory point to mainly additive genetic variance for complex traits},
  author={Hill, William G and Goddard, Michael E and Visscher, Peter M},
  journal={PLoS genetics},
  volume={4},
  number={2},
  pages={e1000008},
  year={2008},
  publisher={Public Library of Science San Francisco, USA}
}

@Manual{Rlanguage,
  title = {R: A Language and Environment for Statistical Computing},
  author = {{R Core Team}},
  organization = {R Foundation for Statistical Computing},
  address = {Vienna, Austria},
  year = {2024},
  url = {https://www.R-project.org/},
}

\end{document}